\documentclass[journal, twocolumn, 10pt, romanappendices,A4paper]{IEEEtran}
\usepackage{indentfirst}
\usepackage{setspace}
\usepackage[numbers,sort&compress]{natbib}
\usepackage{graphicx}
\usepackage{amstext}
\usepackage{algorithm}
\usepackage{algorithmic}
\usepackage{mathrsfs}
\usepackage{amssymb}
\usepackage{amsmath}
\usepackage{epstopdf}
\usepackage{CJK}
\usepackage{multicol}
\usepackage{stfloats}
\usepackage{url}
\usepackage{color}
\usepackage{multirow}
\usepackage{tabularx}
\usepackage{array}

\usepackage{graphicx,color,overpic,psfrag}
\usepackage{times}
\usepackage{latexsym}
\usepackage{bm}
\usepackage{amssymb}
\usepackage{cases}
\usepackage{array}
\usepackage{float}
\usepackage{fancyhdr}
\usepackage{booktabs}

\usepackage{psfrag}
\usepackage{colortbl}
\usepackage[table]{xcolor}
\usepackage{tabu}

\usepackage{enumerate}
\usepackage{subfigure}

\definecolor{mygray}{gray}{.9}
\definecolor{mypink}{rgb}{.99,.91,.95}
\definecolor{mycyan}{cmyk}{.3,0,0,0}

\newcommand{\PreserveBackslash}[1]{\let\temp=\\#1\let\\=\temp}
\newcolumntype{C}[1]{>{\PreserveBackslash\centering}p{#1}}
\newcolumntype{R}[1]{>{\PreserveBackslash\raggedleft}p{#1}}
\newcolumntype{L}[1]{>{\PreserveBackslash\raggedright}p{#1}}

\ifCLASSINFOpdf

\else

\fi

\begin{document}

\title{\Huge{Asymmetrical Uplink and Downlink Transceivers in Massive MIMO Systems}}

\author{Xi Yang,~\IEEEmembership{Member,~IEEE}, Shi Jin,~\IEEEmembership{Senior Member,~IEEE},\\ Geoffrey Ye Li,~\IEEEmembership{Fellow,~IEEE}, and Xiao Li,~\IEEEmembership{Member,~IEEE}

\thanks{Copyright (c) 2015 IEEE. Personal use of this material is permitted. However, permission to use this material for any other purposes must be obtained from the IEEE by sending a request to pubs-permissions@ieee.org.}
\thanks{X. Yang was with the National Mobile Communications Research Laboratory, Southeast University, Nanjing, 210096, P. R. China. She is now with the State Key Laboratory of Internet of Things for Smart City, University of Macau, Taipa, Macao (e-mail: ouyangxi@seu.edu.cn).}
\thanks{S. Jin and X. Li are with the National Mobile Communications Research Laboratory, Southeast University, Nanjing, 210096, P. R. China (e-mail: jinshi@seu.edu.cn; li\_xiao@seu.edu.cn).}
\thanks{G. Y. Li is with the Department of Electrical and Electronic Engineering, Imperial Colledge London, London, UK (e-mail: geoffrey.li@imperial.ac.uk).}
}
\newcommand{\rl}[1]{\color{red}#1}

\maketitle

\begin{abstract}
Even if massive multiple-input multiple-output (MIMO) can theoretically bring huge benefits, it incurs substantial hardware complexity and expensive hardware costs. To address these issues while maintaining the system performance simultaneously, we develop an asymmetrical transceiver architecture for massive MIMO systems in this paper by releasing the shackles on radio frequency (RF) chains. Specifically, we first develop the architecture for the asymmetrical transceiver where the number of receive RF chains is different from that of the transmit RF chains. Due to this unequal number of RF chains, channel inconsistency appears. To overcome the channel inconsistency and thus fully harness the system's downlink data transmission capability, we propose two uplink-to-downlink channel transfer algorithms. The cost and power consumption models for the asymmetrical transceiver are also developed and the uplink signal-to-noise loss due to the channel inconsistency is investigated. Through analyzing system spectral, cost, and energy efficiency, we demonstrate that the proposed asymmetrical transceiver-based massive MIMO system can achieve excellent downlink spectral efficiency while maintaining a reasonable energy efficiency.
\end{abstract}

\begin{IEEEkeywords}
Asymmetrical transceiver, uplink-to-downlink channel transfer, spectral efficiency, energy efficiency.
\end{IEEEkeywords}

\IEEEpeerreviewmaketitle
\newtheorem{Definition}{Definition}
\newtheorem{Lemma}{Lemma}
\newtheorem{Theorem}{Theorem}
\newtheorem{Corollary}{Corollary}
\newtheorem{Proposition}{Proposition}
\newtheorem{Remark}{Remark}

\section{Introduction}

%\IEEEPARstart{T}{his}

By using a large-scale antenna array at the base station (BS), massive multiple-input multiple-output (MIMO) can serve multiple users at the same frequency band simultaneously and thus boost the system spectral efficiency \cite{Marzetta10noncooperative,Larsson14massive,Sun15Beam,Wang18spatial}. Therefore, massive MIMO has become one of the key technologies in the fifth-generation (5G) wireless communication systems and is also anticipated to play an important role in the sixth-generation (6G) wireless communication systems \cite{Zhang196G}.
However, the large-scale antenna array in massive MIMO systems also brings tremendous hardware complexity and power consumption if each radio frequency (RF) chain is connected to one antenna \cite{Shepard13Argos,Yang18Digital}\footnote{Note that in time division duplex-based full digital systems, an independent RF chain pair, which consists of a transmit RF chain and a receive RF chain, is connected to each antenna via a switch.}.
Huge baseband signal processing pressure has also been caused by such a large-scale antenna array and numerous RF chains. As a result, these hardware constraints severely impact the wide deployment of massive MIMO systems \cite{Yang18Digital}.

Recently, several methods have been proposed to overcome these constraints. For example, hybrid (i.e., the analog and digital combined) beamforming \cite{Ayach14Spatially,Liang14Low,Payami16Hybrid,Payami19Phase,Sungwoo17Dynamic} is leveraged in massive MIMO systems to reduce the number of RF chains. Nevertheless, hybrid beamforming usually incurs additional restrictions on the relevant signal processing, such as constant modulus and a heavy beam training overhead for channel estimation.
Low-resolution analog-to-digital converters (ADCs) and digital-to-analog converters (DACs) have also been introduced in \cite{Wang18Finite,Kong17Full,Zhang16ON,Dong20spatially} to reduce the hardware cost and power consumption. {However, in low-resolution massive MIMO systems, part of the signal information is inevitably lost. Therefore, complex iterative algorithms must be used to recover data at the receiver.}
Apart from that, antenna selection \cite{Li14Energy,Rodriguez17Reduced,Gao18MassiveSwitch,Asaad18antennaselection,Chen19Intell} has also been extended to massive MIMO systems to alleviate the requirement on the number of RF chains. Whereas, the RF switching network in such systems becomes less power-efficient due to the increasing MIMO dimension. More importantly, similar to the hybrid beamforming systems, the numbers of transmit and receive RF chains in massive MIMO systems with antenna selection are also reduced, {thereby degrading the data transmission capability at both uplink and downlink.}

In general, the demand on uplink data transmission is different from that of the downlink in practical scenarios. For example, many popular applications receive more data in downlink than in uplink \cite{Ericsson16}. On the other hand, the data transmission rate is highly related to the number of RF chains \cite{Molisch05capacity}. The more RF chains, the higher the transmission rate we can achieve; however, the higher the circuit complexity and cost we need to address. With massive MIMO, the hardware complexity of RF chains has been greatly increased in 5G while the budgets for size, power, and cost for the system nearly remain the same \cite{Shepard20what}.
In addition, from \cite{Yang18Digital}, the cost and power consumption of the massive MIMO transceiver circuit for the fully digital beamforming architecture and the hybrid beamforming architecture are almost the same. The major gap on cost and power consumption is contributed by the ADCs, DACs, and baseband processing field-programmable gate arrays (FPGAs).
{Therefore, an alternatively feasible way to alleviate the cost and hardware challenges is by decoupling the transmit and the receive RF chains so that the numbers of the receive and transmit RF chains can be set different.}

Based on this idea, a full digital system architecture with nonreciprocal antenna arrays has been proposed in \cite{Guo20Design}, {where only a few antennas in the center part of the antenna array at the base station} are connected with both transmit and receive RF chains while the others only connect with transmit RF chains. Owing to this asymmetry in the numbers of transmit and receive antennas, we call such an architecture as the asymmetrical transceiver-based architecture in this paper. Benefited from this flexibly unequal numbers of transmit and receive RF chains, the system cost and hardware complexity of the asymmetrical transceiver-based massive MIMO systems can be substantially decreased and the baseband data processing pressure can be alleviated. Consequently, we can achieve a better energy and cost efficiencies, which are the key performance indicators of 6G \cite{You20Towards}.

Inspired by the nonreciprocal array design proposed in \cite{Guo20Design}, we investigate the detailed uplink and downlink transmission procedures and the system performance of the asymmetrical transceiver-based massive MIMO system in this paper.
In particular, we first provide a generic system architecture for the asymmetrical transceiver, where the number of receive RF chains is fewer than that of transmit RF chains. Then, the overall transmission procedure of the asymmetrical transceiver-based massive MIMO system is presented. Afterwards, we design two uplink-to-downlink channel transfer algorithms. Finally, we develop cost and power consumption models for the asymmetrical transceiver-based massive MIMO system and investigate the spectral efficiency (SE), cost, and energy efficiency (EE) of the asymmetrical transceiver-based system.
The main contributions can be summarized as follows:
\begin{itemize}
\item We provide a general architecture and an overall transmission procedure for the asymmetrical transceiver-based massive MIMO system. The impact of the channel inconsistency due to the unequal number of transmit and receive RF chains is also investigated. Owing to the asymmetrical architecture, the asymmetrical transceiver can significantly reduce the overall system hardware complexity while the signal processing flexibility and superior downlink performance can still be maintained.

\item We propose two uplink-to-downlink channel transfer algorithms, i.e., the discrete Fourier transform (DFT)-based algorithm and the modified Newtonized orthogonal matching pursuit (mNOMP)-based algorithm, to deal with the unequal number of transmit and receive RF chains. {The computational complexity and convergence of the two uplink-to-downlink channel transfer algorithms are also discussed.} Collaborated with the random receive antenna array, these two algorithms can well recover the downlink channel information, which makes the asymmetrical transceiver demonstrate excellent downlink transmission capability.

\item We develop {cost and power consumption models} for the asymmetrical transceiver-based massive MIMO system. The SE and EE comparisons with other conventional symmetrical transceivers are performed as well. Numerical results demonstrate that the proposed asymmetrical transceiver can achieve a superior downlink SE while maintaining a well EE simultaneously. The asymmetrical transceiver architecture is a promising alternative solution for future massive MIMO systems.

\end{itemize}

The rest of this paper is organized as follows. In Section II, we present the system architecture and the channel model for the asymmetrical transceiver-based massive MIMO system. Section III investigates the overall transmission procedure. Two uplink-to-downlink channel transfer algorithms are proposed in Section IV and the cost and EE are analyzed in Section V. Section VI provides the numerical results. We conclude the paper in Section VIII.

\emph{Notation:} Bold lowercase $\mathbf{x}$ and bold uppercase $\mathbf{X}$ are used to denote vectors and matrices, respectively. The inverse, transpose, conjugate and conjugate-transpose operations of matrix are denoted by $(\cdot)^{-1}$, $(\cdot)^{T}$, $(\cdot)^{*}$ and $(\cdot)^{H}$, respectively. $\mathbf{I}_M$ is an identity matrix with dimension $M\times M$. $\left\| \mathbf{x} \right\|$ represents the Euclidean norm of the vector $\mathbf{x}$ and ${\left| \mathcal{A}  \right|}$ stands for the ensemble of the set $\mathcal{A}$.
$[\mathbf{x}]_{m}$ indicates the $m$th element of the vector $\mathbf{x}$ while $[\mathbf{x}]_{\mathcal{A}}$ represents a new vector with $[\mathbf{x}]_{\mathcal{A}}$'s elements coming from $\mathbf{x}$ whose selected elements are specified by the set ${\mathcal{A}}$.

\section{System Model}
In this section, we will first introduce the asymmetrical transceiver-based massive MIMO system and then describe the uplink and downlink channel models for the asymmetrical transceiver.

\subsection{System Architecture}
\begin{figure}[!t]
	\centering
	\includegraphics[scale= 0.15]{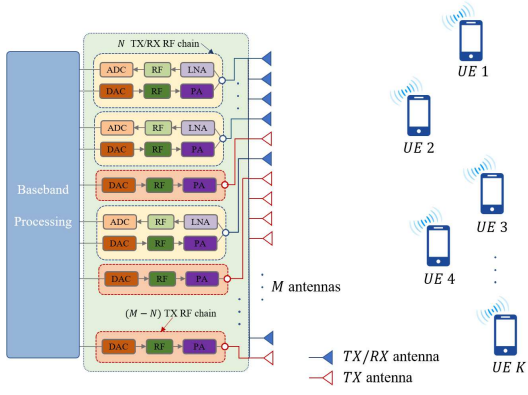}
	\caption{The system architecture of an asymmetrical transceiver-based massive MIMO system, where a BS is equipped with an $M$-element ULA and serves $K$ single-antenna users simultaneously. An asymmetrical transceiver is adopted at the BS with $N$ out of $M$ antennas connecting with both transmit and receive RF chains, while the remaining $M-N$ antennas connect with only transmit RF chains.}\label{Fig:system_architecture}
\end{figure}

The architecture of an asymmetrical transceiver-based multi-user massive MIMO system is shown in Fig.\,\ref{Fig:system_architecture}. Different from the conventional symmetrical transceivers where each transmit RF chain is coupled with a receive RF chain, the BS in Fig.\,\ref{Fig:system_architecture} adopts an asymmetrical transceiver and serves $K$ single-antenna users simultaneously.
For simplicity, we assume that an $M$-element uniform linear array (ULA) is equipped at the BS and the adjacent antenna element spacing is half carrier wavelength.\footnote{The developed hardware architecture can also be easily extended to other array topologies, such as uniform planar antenna array (UPA) and uniform cylinder antenna array (UCA). Note that the subsequent analyses including the system model and the uplink-to-downlink channel transfer algorithms are primarily designed for ULA. When UPA or UCA is employed, the following system model and channel transfer algorithms may necessitate appropriate modifications or reconsideration.} We also assume that this system operates in time-division duplex (TDD) mode and the uplink and downlink transmission can be performed in a coherent interval such that the channel reciprocity between uplink and downlink can be leveraged.

\begin{figure}[!t]
	\centering
	\includegraphics[scale= 0.17]{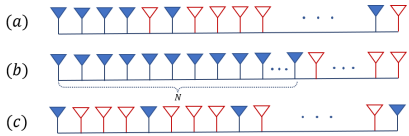}
	\caption{Receive antenna arrays constructed by three receive antenna selection methods: (a) random antenna selection, (b) successive antenna selection, and (c) comb antenna selection. The blue filled one is selected for both transmitting and receiving while the red one is only used for transmitting. The total number of the BS antenna array elements is $M$, and the number of antenna elements utilized for both transmitting and receiving is $N$, which is consistent with Fig.\,\ref{Fig:system_architecture}.}\label{Fig:antenna_selection}
\end{figure}

Note that in an asymmetrical transceiver-based system, the number of receive RF chains is different from that of transmit RF chains, depending on whether we want to boost the system uplink transmission rate or reduce the uplink baseband processing pressure. By considering that the amount of required data is relatively smaller than that of the downlink \cite{ITU17Minimum,3GPP19overall}, we assume the number of the receive RF chains is fewer than that of the transmit RF chains in this paper. As shown in Fig.\,\ref{Fig:system_architecture}, each of the $M$ antennas at the BS connects with a transmit RF chain, but only $N$ ($N \ge K$) out of $M$ antennas are connected with the receive RF chains. In other words, only $N$ antennas connect with both transmit and receive RF chains. Besides, it is worth mentioning that the proposed asymmetrical transceiver-based BS utilizes a full digital beamforming architecture. Therefore, the BS can always employ full digital beamforming/combining at downlink/uplink transmission and thus achieves a higher signal processing flexibility than the hybrid beamforming architecture. The hybrid beamforming architecture usually leverages phase shifters and thereby suffers from unitary modulus constraints. In addition, because of the reduction of both the expensive RF components and the high-rate ADCs, the system cost and hardware complexity of the asymmetrical transceiver-based massive MIMO systems can be greatly reduced and the high uplink baseband data processing pressure can also be alleviated at the cost of moderate uplink transmission capacity reduction.

As is known that the topology of receive antenna array has a great influence on the system performance, thus selecting appropriate antenna elements to connect with those limited number of receive RF chains is important.
Although antenna selection is not a new topic and there are various antenna selection algorithms for antenna selection systems \cite{Asaad18antennaselection,Gao13Antenna,Dua06Receive}, those exiting antenna selection algorithms proposed for symmetrical transceivers cannot be directly applied to asymmetrical transceivers owing to the different system architecture.
Therefore, to uncover the potential performance of asymmetrical transceivers, we propose three intuitive antenna selection methods in this paper\footnote{The sophisticated antenna selection criteria, such as maximizing the angular resolution of the receive antenna array, minimizing the interference signal level from other directional users, and so on, are left for our future research direction.} as shown in Fig.\,\ref{Fig:antenna_selection}, i.e., (a) random antenna selection method \cite{Lo64Mathematical}, (b) successive antenna selection method, and (c) comb antenna selection method.
For the random antenna selection method, the antenna elements used for signal reception are randomly selected; while for the successive selection method, $N$ consecutive antenna elements are selected for signal receiving. In contrast, the receive antenna elements selected by the comb selection method are uniformly distributed across the whole BS antenna array.

\subsection{Channel Model}
We adopt the block-fading parametric channel model\footnote{Although the statistical channel model \cite{Ngo13energy} is widely utilized in wireless communication systems, especially in sub-6G systems, we mainly focus on the parametric channel modeling in this paper. The research considering the statistical channel model is left in our future work and thus is beyond the scope of this paper.} in this paper. Due to the channel reciprocity\footnote{In this paper, we assume that the hardware components are ideally calibrated such that we can leverage the channel reciprocity and ignore the hardware mismatch between the transmit RF chains and the receive RF chains.}, the uplink and downlink channel of user $k$ in the asymmetrical transceiver-based multi-user massive MIMO system can be expressed as \cite{Yin13acoordinated,Wang19anoverview}
\begin{equation}\label{eq:hku}
{\bf{h}}_{k,{\rm{A}}}^{\rm{U}} = \sqrt {\frac{N}{P_k}} \sum\limits_{i = 1}^{P_k} {{g_{k,i}}{{\bf{a}}_{\rm{U}}}({\theta _{k,i}})},
\end{equation}
and
\begin{equation}\label{eq:hkd}
{\mathbf{h}}_{k}^{\rm{D}} = \sqrt {\frac{M}{P_k}} \sum\limits_{i = 1}^{P_k} {{{g}_{k,i}}{{\mathbf{a}}_{\rm{D}}}({{\theta }_{k,i}})},
\end{equation}
respectively, where $P_k$ denotes the number of paths; ${g_{k,i}}$ is the complex gain of the $i$th path for the $k$th user; $\theta_{k,i}$ represents the angle of arrival/departure (AoA/AoD) of the $i$th path at uplink/downlink. ${{\bf{a}}_{\rm{U}}}(\cdot)\in\mathbb{C}^{N\times 1}$ and ${{\mathbf{a}}_{\rm{D}}}(\cdot)\in\mathbb{C}^{1\times M}$ are the uplink and downlink steering vectors, respectively, and can be expressed as
\begin{equation}\label{eq:steeringVectorau}
{{\bf{a}}_{\rm{U}}}({\theta _{k,i}}) = \frac{1}{{\sqrt N }}{\left[ {{e^{ - \frac{{j2\pi {a_1}d\sin \theta_{k,i} }}{\lambda }}}, \ldots ,{e^{ - \frac{{j2\pi {a_N}d\sin \theta_{k,i} }}{\lambda }}}} \right]^T},
\end{equation}
and
\begin{equation}\label{eq:steeringVectorad}
{{\bf{a}}_{\rm{D}}}({\theta _{k,i}}) = \frac{1}{{\sqrt M }}{\left[1,{{e^{ - \frac{{j2\pi d\sin \theta_{k,i} }}{\lambda }}}, \ldots ,{e^{ - \frac{{j2\pi (M-1)d\sin \theta_{k,i} }}{\lambda }}}} \right]},
\end{equation}
where $\lambda$ is the carrier wavelength and $d$ is the adjacent antenna element spacing of the original BS antenna array (i.e., the $M$-antenna ULA). Denote ${a_n}\in\{1,2,\ldots,M\},\forall n=1,\ldots,N,$ as the index of the BS's antennas selected for uplink receiving and ${\cal A} \buildrel \Delta \over = \left\{ {{a_1},{a_2}, \ldots {a_N}} \right\}$ as the receive antenna index set.

Due to the unequal numbers of transmit and receive antennas in the asymmetrical transceiver, the expression form of the steering vector in the uplink i.e., (\ref{eq:steeringVectorau}), is different from that in the downlink i.e., (\ref{eq:steeringVectorad}), even for the same user. Note that the steering vectors at uplink and downlink remain the same form for the same user at conventional symmetrical transceiver-based massive MIMO systems.
The following remark is provided to highlight this different feature.
\begin{Remark}\label{Remark_1}
Unlike the existing massive MIMO systems where symmetrical transceivers are generally employed, the channels at the uplink and the downlink in the asymmetrical transceiver-based massive MIMO system are not consistent.
\end{Remark}

In particular, the channel inconsistency between the uplink and the downlink mainly exhibits in two respects. First, the dimensions of the uplink and downlink channel vectors are unequal.
For example, in Fig.\,\ref{Fig:system_architecture}, the dimension of the uplink channel vectors is $N$ while the dimension of the downlink channel vector is $M$ ($M>N$). Owing to this unequal channel vector dimension, the uplink channel estimate cannot be directly used for the downlink precoding. Additional signal processing, called uplink-to-downlink channel transfer, is needed before performing the downlink precoding.

Second, although there are totally $P_k$ paths for user $k$ in the propagation environment, the number of paths the BS can distinguish for user $k$ at uplink and downlink may be different.
This is primarily incurred by the different antenna array apertures the BS used for uplink and downlink.
Thanks to the asymmetrical transceiver architecture, the transmit antenna array (i.e., the $M$-antenna ULA) of the BS always has a larger aperture and thus acquires a better angular resolution than the receive antenna array (i.e., the selected $N$-antenna array).
For example, for the successive antenna selection under our system configuration, we have the angular resolution of the receive antenna array $\Delta^s_r=2/N$, while the angular resolution of the transmit antenna array is $\Delta_t=2/M$ and $\Delta^s_r>\Delta_t$.
Hence, if the AoA difference of the $i$th path and the $j$th path $\Delta_{\theta_{ij}}$ is smaller than $\Delta^s_r$ but larger than $\Delta_t$, i.e., $\Delta_t<\Delta_{\theta_{ij}}<\Delta^s_r$, the $i$th path and the $j$th path can be distinguished at the BS in downlink but may not be distinguished in uplink.
In this situation, the uplink channel of user $k$ from the view of BS becomes
\begin{equation}\label{eq:hkus}
{\bf{h}}_{k,{\rm{A}}}^{\rm{U,s}} = \sqrt {\frac{N}{P_k}} \sum\limits_{i = 1}^{P^s_k} {{g^{{s}}_{k,i}}{{\bf{a}}^s_{\rm{U}}}({\theta^s_{k,i}})},
\end{equation}
where ${g^{{s}}_{k,i}}$ and ${\theta^s_{k,i}}$ denote the complex gain and the AoA of the $i$th identified path, respectively, from the composition of the unresolvable paths, and the steering vector of the selected antenna array can be written as
\begin{equation}\label{eq:steeringVectoraus}
{{\bf{a}}^s_{\rm{U}}}({\theta^s_{k,i}}) = \frac{1}{{\sqrt N }}{\left[1,{{e^{ - \frac{{j2\pi d\sin {\theta^s_{k,i}} }}{\lambda }}}, \ldots ,{e^{ - \frac{{j2\pi (N-1)d\sin {\theta^s_{k,i}} }}{\lambda }}}} \right]},
\end{equation}
$P^s_k(\leq P_k)$ represents the actual number of paths that the BS can resolve.

\begin{figure}[!t]
	\centering
	\includegraphics[scale= 0.4]{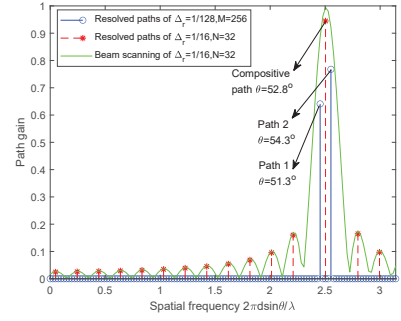}
	\caption{An example for the channel inconsistency at the path number/angle in asymmetrical transceiver-based massive MIMO systems. Because of $M=256$ and $N=32$, we have $\Delta_t=0.45^o$ and $\Delta^s_r=3.58^o$. Therefore, those two paths are well resolved at downlink but only one dominant path whose compositive angle is $52.8^o$ can be resolved at uplink.}\label{Fig:unresolved_paths}
\end{figure}

To better illustrate this phenomenon, Fig.\,\ref{Fig:unresolved_paths} provides a detailed example. We assume that there are two dominant paths existing in the wireless propagation environment, whose angles are $51.3^o$ and $54.3^o$, respectively. At the BS, a ULA is used and the number of transmit antennas is $M=256$ (i.e., $\Delta_t=0.45^o$). For the receive antenna array at the BS, we employ the successive antenna selection and assume $N=32$ (i.e., $\Delta^s_r=3.58^o$). As presented in Fig.\,\ref{Fig:unresolved_paths}, since $0.45^o<3^o<3.58^o$, those two paths can be well resolved at downlink while only one dominant path whose angle is $52.8^o$ can be resolved at uplink.
Note that the best angular resolution of the receive antenna array in (\ref{eq:hku}) can be obtained as $\Delta_r=\Delta_t$ via the appropriate antenna selection.
Therefore, despite the unequal channel vector dimension, the unequal angular resolutions of the receive and the transmit antenna arrays should also be considered. The receive antenna array topology design at the BS in terms of the effective array aperture is nontrivial for the asymmetrical transceiver-based massive MIMO systems.

\section{Transmission Procedure}
In this section, we illustrate the overall transmission procedure of the asymmetrical transceiver-based massive MIMO system, including the uplink transmission, the uplink-to-downlink channel transfer, and the downlink transmission.

\subsection{Overall Transmission Procedure}
\begin{figure}[!t]
	\centering
	\includegraphics[scale= 0.17]{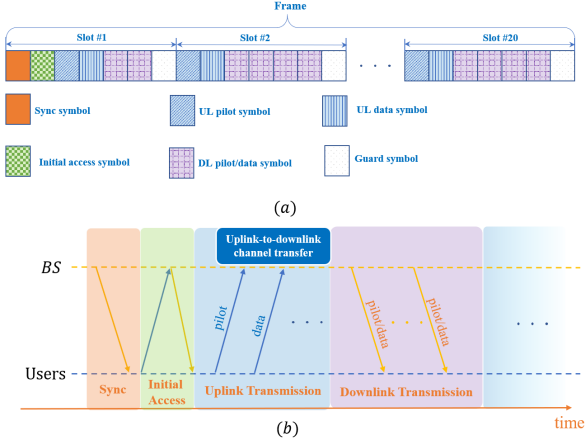}
	\caption{The wireless frame structure and the transmission procedure of the asymmetrical transceiver-based massive MIMO system.}\label{Fig:frame}
\end{figure}

Fig.\,\ref{Fig:frame}(a) presents the frame structure of the asymmetrical transceiver-based massive MIMO system. Since the system operates in TDD mode, there is a guard symbol when switching from the downlink to the uplink and the transmission procedure is demonstrated in Fig.\,\ref{Fig:frame}(b).

During the synchronization and initial access symbols, the system synchronization and the user initial access are performed. First, the BS broadcasts the synchronization sequences to the users so that they can accomplish their frequency and time synchronization upon the received sequences.
After that, users start the initial access and then move into the uplink transmission when they acquire the acknowledgement from the BS. During the uplink transmission period, uplink pilots are first sent by the $K$ users to perform uplink channel estimation.
Since the full digital architecture is employed for data reception, the BS can theoretically estimate the uplink channels for $K$ users after $K$ measurements. This significantly reduces the training overhead when compared with the hybrid architecture. For the conventional hybrid transceiver, the number of channel estimation measurements is usually proportional to the product of the number of BS's receive antennas and the number of users \cite{heath2016overview}. Moreover, due to the employed much fewer receive RF chains in the asymmetrical transceiver, the uplink baseband data throughput and processing pressure are also considerably reduced at the same time.

After the uplink training, uplink data transmission begins and the acquired uplink channel estimates are used to recover the uplink data. In the conventional symmetrical transceivers, the uplink channel estimates are simultaneously used for the downlink precoding matrix calculation. However, in the asymmetrical transceiver, this cannot be directly performed due to the aforementioned channel inconsistency. To overcome this channel inconsistency, we introduce an additional step, i.e., the uplink-to-downlink channel transfer, before the downlink transmission. In the uplink-to-downlink channel transfer, the uplink channel estimates are extended and refined to obtain the downlink channel estimates. Then, the downlink precoding is accomplished based on these recovered downlink channel estimates and the downlink pilot/data transmission is subsequently performed.
In the following subsections, we present the details of the transmission procedure.

\subsection{Uplink Transmission}
The uplink transmission includes the uplink pilot training and the uplink data transmission.
During the uplink pilot training, we assume that users transmit orthogonal pilot sequences. Denote the length of the pilots as $\tau(\tau\geq K)$, we have
\begin{equation}\label{eq:Y}
{{\bf{Y}} } = \sqrt {{\rho_\tau }} {\bf{H}}_{\rm{A}}^{\rm{U}}{{\bf{P}} } + {\bf{N}},
\end{equation}
where ${{\bf{Y}} } \in {\mathbb{C}^{N \times \tau }}$ is the received signal at the BS, $\rho_\tau$ denotes the pilot transmit power by each user, ${{\bf{P}} } \in {\mathbb{C}^{K \times \tau }}$ is the normalized pilot matrix, and ${\bf{N}} \in {\mathbb{C}^{N \times \tau }}$ represents an additive white, zero-mean complex Gaussian noise matrix with element's variance being $\sigma_n^2=1$ without loss of generality. ${\bf{H}}_{\rm{A}}^{\rm{U}}\in {\mathbb{C}^{N \times K}}$ denotes the composite uplink channel matrix and can be expressed as ${\bf{H}}_{\rm{A}}^{\rm{U}} = \left[ {{\bf{h}}_{1{\rm{,A}}}^{\rm{U}}, \cdots ,{\bf{h}}_{K{\rm{,A}}}^{\rm{U}}} \right]$.

When the least-square (LS) channel estimation is utilized, we obtain
\begin{equation}\label{eq:HAuestimatedLS}
{\bf{\hat H}}_{\rm{A}}^{\rm{U}} = \frac{1}{{\sqrt {{\rho_\tau }} }}{\bf{Y}}{\bf{P}}^H,
\end{equation}
where the columns of ${\bf{\hat H}}_{\rm{A}}^{\rm{U}}$ denote the uplink channel estimates for different users, i.e., ${\bf{\hat H}}_{\rm{A}}^{\rm{U}} = \left[ {{\bf{\hat h}}_{1{\rm{,A}}}^{\rm{U}}, \cdots ,{\bf{\hat h}}_{K{\rm{,A}}}^{\rm{U}}} \right]$.

When the linear minimum mean-squared error (LMMSE) channel estimation is employed, the uplink channel can be estimated as
\begin{equation}\label{eq:HAuestimatedLMMSE2}
{\bf{\bar H}}_{\rm{A}}^{\rm{U}} = \frac{1}{{\sqrt {{\rho _\tau }} }}{\bf{Y}}{{\bf{P}}^H}{\left( {\frac{1}{{{\rho _\tau }}}{\bf{R}}_{{\bf{H}}_{\rm{A}}^{\rm{U}}}^{ - 1} + {{\bf{I}}_K}} \right)^{ - 1}},
\end{equation}
where ${{\bf{R}}_{{{\bf{H}}^{\rm{U}}_{\rm{A}}}}} \buildrel \Delta \over = \mathbb{E}\left\{ {{{\left( {{\bf{H}}_{\rm{A}}^{\rm{U}}} \right)}^H}{\bf{H}}_{\rm{A}}^{\rm{U}}} \right\}$ with ${\bf{\bar H}}_{\rm{A}}^{\rm{U}} = \left[ {{\bf{\bar h}}_{1{\rm{,A}}}^{\rm{U}}, \cdots ,{\bf{\bar h}}_{K{\rm{,A}}}^{\rm{U}}} \right]$ is the correlation matrix of the uplink channel.

After the uplink pilot training, uplink data is transmitted. Denote ${\bf{x}} = {\left[ {{x_1}, \ldots ,{x_K}} \right]^T}$ as the transmitted uplink data, where $x_k$ represents the symbol transmitted by the $k$th user and $\mathbb{E}\left\{ {{{\left| {{x_k}} \right|}^{{2}}}} \right\} = 1,\forall k = 1, \ldots ,K$, then, we have
\begin{equation}\label{eq:rMRC}
{{\bf{r}}_{{\rm{MRC}}}} = \sqrt {{\rho_u}} {\left( {{\bf{\tilde H}}_{\rm{A}}^{\rm{U}}} \right)^H}{\bf{H}}_{\rm{A}}^{\rm{U}}{\bf{x}} + {\left( {{\bf{\tilde H}}_{\rm{A}}^{\rm{U}}} \right)^H}{\bf{n}}_u,
\end{equation}
for the MRC receiver\footnote{To simplify the analysis and better highlight the insights on how the system parameters impact the uplink ergodic achievable SE, we have adopted the MRC detector here. Notice that other detectors, such as zero-forcing and LMMSE detectors, can also be utilized for uplink data transmission.}, where ${{\bf{r}}_{{\rm{MRC}}}} \in {\mathbb{C}^{K \times {{1}}}}$ is the recovered signal vector, $\rho_u$ represents the transmit data power by each user, and ${\bf{n}}_u\sim\mathcal{CN}({\bf{0}},{{\bf{I}}_N})$ denotes the additive white Gaussian noise. ${\bf{\tilde H}}_{\rm{A}}^{\rm{U}} = \left[ {{\bf{\tilde h}}_{1{\rm{,A}}}^{\rm{U}}, \cdots ,{\bf{\tilde h}}_{K{\rm{,A}}}^{\rm{U}}} \right]$ is the channel estimate from the uplink pilot training. When the LS channel estimation is adopted, ${\bf{\tilde H}}_{\rm{A}}^{\rm{U}} = {\bf{\hat H}}_{\rm{A}}^{\rm{U}}$, and when the LMMSE channel estimation is adopted, we have ${\bf{\tilde H}}_{\rm{A}}^{\rm{U}} = {\bf{\bar H}}_{\rm{A}}^{\rm{U}}$.
Therefore, the ergodic uplink achievable SE of the $k$th user under maximum ratio combining (MRC) can be expressed as
\begin{align}\label{eq:rMRCkthSE}
&R_{{\rm{MRC}},k}^{\rm{U}}= \nonumber \\
&\mathbb{E} \left\{ {{{\log }_2}\left( {1 + \frac{{{\rho_u}{{\left| {{{\left( {{\bf{\tilde h}}_{k,{\rm{A}}}^{\rm{U}}} \right)}^H}{\bf{h}}_{k,{\rm{A}}}^{\rm{U}}} \right|}^2}}}{{{\rho_u}\sum\nolimits_{i = 1,i \ne k}^K {{{\left| {{{\left( {{\bf{\tilde h}}_{k,{\rm{A}}}^{\rm{U}}} \right)}^H}{\bf{h}}_{i,{\rm{A}}}^{\rm{U}}} \right|}^2} + {{\left\| {{\bf{\tilde h}}_{k,{\rm{A}}}^{\rm{U}}} \right\|}^2}} }}} \right)} \right\},
\end{align}
and the ergodic uplink system achievable SE can be given by
\begin{equation}\label{eq:rMRCsysSE}
\eta _{{\rm{SE}}}^{\rm{U}} = \sum\limits_{i = 1}^K {R_{{\rm{MRC}},i}^{\rm{U}}}\quad{\rm{ (bit/s/Hz)}}.
\end{equation}

From (\ref{eq:hku}), (\ref{eq:steeringVectorau}), and (\ref{eq:hkus}), the ergodic uplink SE, i.e., $R_{{\rm{MRC}},k}^{\rm{U}}$ in (\ref{eq:rMRCkthSE}), is affected by both the antenna number and the topology of the receive antenna array.
Since the angular resolution of the receive antenna array from the comb antenna selection is $\Delta^c_r=2/M$, which is equal to $\Delta_t=2/M$, the channel model used in the analysis of $R_{{\rm{MRC}},k}^{\rm{U}}$ is ${\bf{h}}_{k,{\rm{A}}}^{\rm{U}}$ in (\ref{eq:hku}). While for the successive and the random antenna selections, the angular resolution of the realized receive antenna array is either proportional to the number of antenna elements or to the aperture dimension the selected elements spread. Therefore, the impact of the angular mismatch due to the degraded angular resolution should be considered in analyzing $R_{{\rm{MRC}},k}^{\rm{U}}$.
Define the normalized uplink SNR loss as
\begin{equation}
{\rm{SNR}_{{\rm{loss}}}}\buildrel \Delta \over =1-\frac {{{{\left| {{{\left(  {{\bf{\tilde h}}_{k,{\rm{A}}}^{\rm{U}}} \right)}^H}{\bf{h}}_{k,{\rm{A}}}^{\rm{U}}} \right|}^2}}}{{ {{\left\|  {{\bf{\tilde h}}_{k,{\rm{A}}}^{\rm{U}}} \right\|}^2}}{ {{\left\|  {\bf{h}}_{k,{\rm{A}}}^{\rm{U}} \right\|}^2}} }.
\end{equation}
We can find the uplink SNR loss due to the degraded angular resolution under single-user scenarios in the following.
\begin{Proposition}\label{Pro1}
Considering the degraded angular resolution from successive antenna selection, when the two resolvable equal-power paths\footnote{Since these two paths come from adjacent angles, we assume the power of these two paths are equal for simplicity. However, the following analysis can be easily extended to unequal-power paths.} (i.e., $P_k=2$, ${\Delta_r=\Delta_t<|\theta_{k,1}-\theta_{k,2}| \leq \Delta^s_r}$) can only be resolved as one dominant path at BS (i.e., ${P_k^s=1}$ with the compositive angle being $\theta^s_{k,1}$), the normalized uplink SNR loss of user $k$ with no channel estimation error from white noise can be calculated as
\begin{align}
{\rm{SNR}_{{\rm{loss}}}} & = 1 - \frac{{\Lambda _1^2 + \Lambda _2^2 + 2{\Lambda _1}{\Lambda _2}\cos \Gamma}}{{2{N^2} + 2N\Lambda \cos \Gamma}}, \label{eq:SNRloss}
\end{align}
where ${\Lambda _1} \buildrel \Delta \over = \frac{{\sin \left( {\pi dN{\Theta _1}/\lambda } \right)}}{{\sin \left( {\pi d{\Theta _1}/\lambda } \right)}}$, ${\Lambda _2} \buildrel \Delta \over = \frac{{\sin \left( {\pi dN{\Theta _2}/\lambda } \right)}}{{\sin \left( {\pi d{\Theta _2}/\lambda } \right)}}$, $\Lambda  \buildrel \Delta \over = \frac{{\sin \left( {\pi dN\Theta /\lambda } \right)}}{{\sin \left( {\pi d\Theta /\lambda } \right)}}$, $\Theta  \buildrel \Delta \over = \sin {\theta _{k,1}} - \sin {\theta _{k,2}}$, ${\Theta _1} \buildrel \Delta \over = \sin \theta _{k,1}^s - \sin {\theta _{k,1}}$, ${\Theta _2} \buildrel \Delta \over = \sin \theta _{k,1}^s - \sin {\theta _{k,2}}$, $\Gamma  \buildrel \Delta \over = {\phi _1} - {\phi _2} - \pi d(N - 1)\Theta /\lambda$, $\phi_1$ and $\phi_2$ are the complex gains' phases of the original two paths, and we have
\begin{align}
0< {\rm{SNR}_{{\rm{loss}}}} < 1. \label{eq:SNRlossrange}
\end{align}
\end{Proposition}
\begin{IEEEproof}
See Appendix A.
\end{IEEEproof}

From (\ref{eq:SNRloss}), the normalized uplink SNR loss is not only related to the number of receive antennas and the angle difference between the two paths, but also related to the phase difference between the two paths' complex gains. With the fixed number of receive antennas and angle difference and $\Theta_1=\Theta_2$, the normalized uplink SNR loss decreases with the increase of the cosine function of the phase difference, i.e., $\cos \Gamma$. It should also be mentioned that when $\cos\Gamma<0$, there is a high probability that two paths can be still resolved even if $|\theta_{k,1}-\theta_{k,2}| < \Delta^s_r$. Interestingly, due to such destructive superposition, the angle difference of these resolved two effective paths will be larger than the angular resolution of the receive antenna array, i.e., $|\theta_{k,1}'-\theta_{k,2}'| >\Delta^s_r$. The numerical results in Section VI will demonstrate this phenomenon.

Additionally, thanks to the accomplished different angular resolutions by different antenna selection methods, the numerical results presented later also indicate that the random antenna selection outperforms the successive antenna selection in terms of the ergodic uplink system SE.
This means that uplink transmission can benefit from the flexible receive antenna topology design in the asymmetrical transceiver-based massive MIMO systems. Furthermore, although the comb antenna selection also achieves a superior angular resolution as the random antenna selection, it suffers from angle ambiguity incurred by the grating lobe, which consequently degrades the channel estimation performance and the achievable SE.

\subsection{Uplink-to-Downlink Channel Transfer}
Due to the difference between the numbers of receive and transmit antennas at the BS, the uplink channel information cannot be directly utilized for downlink data precoding even in the TDD mode and with physical channel reciprocity. Therefore, it is necessary to figure out how to deduce the downlink channel information from the uplink channel estimates in asymmetrical transceivers.

According to the reciprocity between the uplink and downlink channels in TDD systems, we have
\begin{equation}\label{eq:hdl}
{\left[ {{\mathbf{h}}_k^{\rm D}} \right]_{\mathcal{A}}} = {\left( {{\mathbf{h}}_{k,{\rm A}}^{\rm U}} \right)^T},
\end{equation}
where ${\mathbf{h}}_k^{\rm D} \in {\mathbb{C}^{1 \times M}}$ is the downlink channel of user $k$.
Since $|\mathcal{A}|=N$ and ${\mathbf{h}}_{k,{\rm A}}^{\rm U} \in {\mathbb{C}^{N \times 1}}$, ${\mathbf{h}}_{k,{\rm A}}^{\rm U}$ only contains $N$ elements of ${\mathbf{h}}_k^{\rm D}$. Hence, $M-N$ elements in ${\mathbf{h}}_k^{\rm D}$ need to be recovered.
Unfortunately, in the asymmetrical system, the effective aperture of the uplink receive antenna array may be smaller than that of the downlink transmit antenna array. Thus, the angular resolution of the transceiver at the uplink is worse than that of the downlink. Hence, to effectively acquire the large transmit antenna array gain at downlink, it is essential for the uplink-to-downlink channel transfer algorithms to not only fully exploit the channel information, e.g., AoA, based on the received small-dimension signals at the uplink, but also use this uplink channel information to recover and refine the channel estimates for the downlink.\footnote{ Apart from the uplink-to-downlink channel transfer algorithms, a subarray-based switch network collaborated with a round-robin scheme can be also introduced in the asymmetrical transceiver-based BS to acquire the whole channel information. However, undesired insertion loss will also be brought simultaneously, and the round-robin scheme will result in additional pilot training overhead, which makes the system inefficient and may become unacceptable when the number of receive RF chains is relatively small.}

Substituting (\ref{eq:hku}) into (\ref{eq:hdl}), we obtain
\begin{equation}\label{eq:hdl2}
{\left[ {{\mathbf{h}}_k^{\rm D}} \right]_{\mathcal{A}}} = \sqrt {\frac{N}{P_k}} \sum\limits_{i = 1}^{P_k} {{g_{k,i}}{{\bf{a}}^T_{\rm{U}}}({\theta _{k,i}})}.
\end{equation}
Since there is a known relative topology shared by the transmit and receive antenna array, i.e., the relationship between ${{\bf{a}}_{\rm{U}}}({\theta _{k,i}})\in\mathbb{C}^{N\times 1}$ and ${{\bf{a}}_{\rm{D}}}({\theta _{k,i}})\in\mathbb{C}^{1\times M}$, we only need to determine $P_k$, $g_{k,i}$, and $\theta_{k,i}$ to recover user $k$'s downlink channel. Note that we can merely obtain the uplink channel estimate ${\mathbf{\tilde h}}_{k,{\rm A}}^{\rm U}$ instead of the perfect uplink channel ${{\mathbf{h}}_{k,{\rm A}}^{\rm U}}$. Then, the objective of the uplink-to-downlink channel transfer algorithms has been transformed as recovering ${\mathbf{h}}_k^{\rm D}$ from ${\mathbf{\tilde h}}_{k,{\rm A}}^{\rm U}$.
In order to acquire the potential downlink performance of the asymmetrical transceivers, we exploit uplink-to-downlink channel transfer algorithms in Section IV.

\subsection{Downlink Transmission}
When the downlink channel information is obtained via the uplink-to-downlink channel transfer algorithms, downlink pilot/data transmission begins.
Assume the obtained downlink channel estimate of the $k$th user from the uplink-to-downlink channel transfer algorithms is ${\mathbf{\tilde h}}_k^{\rm D} \in {\mathbb{C}^{1 \times M}}$, then, the downlink channel estimation matrix for these $K$ users can be written as
\begin{equation}\label{eq:HdownlinkK}
{\mathbf{\tilde H}}_{\text{A}}^{\text{D}} = {\left[ {{{\left( {{\mathbf{\tilde h}}_1^{\text{D}}} \right)}^T}, \cdots ,{{\left( {{\mathbf{\tilde h}}_K^{\text{D}}} \right)}^T}} \right]^T} \in {\mathbb{C}^{K \times M}}.
\end{equation}
Let ${\mathbf{s}} = {\left[ {{s_1}, \ldots ,{s_K}} \right]^T}$ be the signal vector transmitted by the BS to these $K$ single-antenna users and $\mathbb{E}\{ {{{\left| {{s_k}} \right|}^{{2}}}} \}{{ = 1,}}\forall k = 1, \ldots ,K$, then the received signal vector by users can be expressed as
\begin{equation}
{\mathbf{v}} = \sqrt {{\rho_d}} {\mathbf{H}}_{\text{A}}^{\text{D}}{\mathbf{Ws}} + {\mathbf{n}}_d\in {\mathbb{C}^{K \times 1}},
\end{equation}
where $\rho_d$ is the downlink transmit power for each user; ${\mathbf{H}}_{\text{A}}^{\text{D}} = {\left[ {{{\left( {{\mathbf{h}}_1^{\text{D}}} \right)}^T}, \cdots ,{{\left( {{\mathbf{h}}_K^{\text{D}}} \right)}^T}} \right]^T} \in {\mathbb{C}^{K \times M}}$ denotes the composite downlink channel; ${\mathbf{W}} = \left[ {{{\mathbf{w}}_1}, \ldots ,{{\mathbf{w}}_K}} \right] \in {\mathbb{C}^{M \times K}}$ is the column-normalized downlink precoding matrix, whose detailed expression is based on ${\mathbf{\tilde H}}_{\text{A}}^{\text{D}}$ and the precoding scheme we selected; ${\mathbf{n}}_d \in {\mathbb{C}^{K \times 1}}$ denotes the additive Gaussian white noise with each of its element satisfying $n_{d,k}\sim\mathcal{CN}(0,1)$. Hence, the received signal at the $k$th user can be expressed as
\begin{equation}
{v_k}  = \sqrt {{\rho_d}} {\mathbf{h}}_k^{\text{D}}{{\mathbf{w}}_k}{s_k} + \sum\limits_{i = 1,i \ne k}^K {\sqrt {{\rho_d}} {\mathbf{h}}_k^{\text{D}}{{\mathbf{w}}_i}{s_i}}  + {n_{d,k}},
\end{equation}
and the ergodic downlink SE of the $k$th user can be given by
\begin{equation}\label{eq:HdownlinkRk}
R_k^{\rm D} = \mathbb{E}\left\{ {{{\log }_2}\left( {1 + \frac{{{{\left| {{\mathbf{h}}_k^{\text{D}}{{\mathbf{w}}_k}} \right|}^2}}}{{\sum\nolimits_{i = 1,i \ne k}^K {{{\left| {{\mathbf{h}}_k^{\text{D}}{{\mathbf{w}}_i}} \right|}^2} + {1 \mathord{\left/
 {\vphantom {1 {{\rho_d}}}} \right.
 \kern-\nulldelimiterspace} {{\rho_d}}}} }}} \right)} \right\}.
\end{equation}
When the maximal-ratio transmitting (MRT) precoding is employed, we have
\begin{equation}\label{eq:Hdownlinkwk}
{{\mathbf{w}}_k} = {{\left( {{\mathbf{\tilde h}}_k^{\rm D}} \right)}^H}/{\xi _k},
\end{equation}
where ${\xi _k} = \| {{\mathbf{\tilde h}}_k^{\rm D}} \|$ is utilized for normalization.
Thus, the downlink system SE is
\begin{equation}\label{eq:HdownlinksysSE}
\eta _{{\text{SE}}}^{\text{D}} = \sum\limits_{k = 1}^K {R_{k}^{\text{D}}} {\text{ (bit/s/Hz)}}.
\end{equation}

From (\ref{eq:HdownlinkRk}) and (\ref{eq:Hdownlinkwk}), the downlink SE of the asymmetrical transceiver-based system is similar to that of the conventional symmetrical transceiver-based full digital massive MIMO system, except for the downlink channel estimate ${{\mathbf{\tilde h}}_k^D}$, where the former endures a signal dimension extension (i.e., the uplink-to-downlink channel transfer) while the latter directly comes from the uplink channel estimates.\footnote{We assume the conventional symmetrical transceiver-based full digital massive MIMO system also operates in TDD mode, and the uplink-to-downlink channel reciprocity maintain perfectly.} Therefore, the uplink-to-downlink channel transfer algorithms, as well as the receive antenna array topology, have a fundamental impact on the downlink performance of the asymmetrical transceiver-based system.

\section{Uplink-to-Downlink Channel Transfer}

In this section, we propose two algorithms for the uplink-to-downlink channel transfer, i.e., the DFT-based channel transfer algorithm and the mNOMP-based channel transfer algorithm. The DFT-based algorithm is mainly designed for the line-of-sight (LOS) path-dominant scenarios for fast channel transfer while the mNOMP-based algorithm is more suitable for multi-path scenarios.

\subsection{DFT-based Channel Transfer}
Based on (\ref{eq:hkd}) and (\ref{eq:steeringVectorad}), the steering vector of paths in ${\left( {{\mathbf{h}}_k^{\rm D}} \right)^T} \in {\mathbb{C}^{M \times 1}}$ asymptotically becomes orthogonal with the columns of the $M$-dimensional DFT matrix when the number of BS antennas continuously increases due to the employment of ULA. Hence, DFT matrix can be used as a spatial matched filter to perform the fast path detection. The underlying principle of the DFT-based channel transfer algorithm is that we first extends the uplink channel estimate ${\mathbf{\tilde h}}_{k,{\rm A}}^{\rm U}$ to an $M$-dimension vector according to the array topology and then executes the spatial matched filtering to recover the downlink channel information. As shown in {\bf Algorithm\,\ref{Alogrithm1}}, the proposed DFT-based channel transfer algorithm has the following steps.
\begin{enumerate}[(i)]
\item {\em Dimension extension}: Pad ${\mathbf{\tilde h}}_{k,{\rm A}}^{\rm U} \in {\mathbb{C}^{N \times 1}}$ with $M-N$ zeros to obtain an $M-$dimensional vector ${\mathbf{\tilde h}}_{k,{\rm S}}^{\rm U}$, i.e.,
\begin{equation}\label{eq:husmissingelement}
{\left[ {{\mathbf{\tilde h}}_{k,{\rm S}}^{\rm U}} \right]_{\mathcal{A}}} = {\mathbf{\tilde h}}_{k,{\rm A}}^{\rm U};
\end{equation}
\item {\em DFT matrix construction}: Set the oversampling factor as $\zeta$ and construct an $M\zeta-$dimensional DFT matrix ${{\mathbf{F}}_{M\zeta }}$ with its $m$th row and $n$th column element being
\begin{equation}
{\left[ {{{\mathbf{F}}_{M\zeta }}} \right]_{mn}} = {e^{ - j2\pi (m - 1)(n - 1)/M\zeta }};
\end{equation}
\item {\em Spatial matched filtering}: Utilize ${{\mathbf{F}}_{M\zeta}}$ to perform the spatial matched filtering and obtain the complex path gains, i.e.,
\begin{equation}\label{eq:dftcalculate}
{{\bf{g}}_{{\rm{DFT}}}} = \frac{{1 }}{ N}{{\bf{F}}_{M\zeta }}{\left[ {{{\left( {{\bf{\tilde h}}_{k,{\rm{S}}}^{\rm{U}}} \right)}^H},{\bf{0}}_{M(\zeta  - 1)}^T} \right]^T}.
\end{equation}
\item {\em Path detection}: Denote ${\tilde g_{[1]}}, \cdots ,{\tilde g_{[N_{\rm peak}]}}$ as the peak elements\footnote{A peak element is a element belonging to the element sequence whose absolute value is larger than its two neighboring elements, i.e., for ${\tilde g_{[i]}},i=1,\ldots,N_{\rm peak}$, we have $|{\tilde g_{[i]}}|\ge |{\tilde g_{[i]-1}}|$ and $|{\tilde g_{[i]}}|\ge |{\tilde g_{[i]+1}}|$.} of the elements of ${{\mathbf{g}}_{\rm{DFT}}}$ and $|{\tilde g_{[1]}}|\ge \cdots \ge|{\tilde g_{[N_{\rm peak}]}}|$.
Then, the number of paths $P_k$, the complex path gains $g_{k,i}$, and the AoAs $\theta_{k,i}$ for user $k$ can be estimated as
\begin{equation}\label{eq:pDft}
\hat P_k=\mathop {\min }\limits_{\hat P_k}\left\{{\left\| {{\bf{\tilde h}}_{k,{\rm{A}}}^{\rm{U}}} \right\|^2} - \sum\limits_{i = 1}^{{{\hat P}_k}} {{{N\left| {{{\tilde g}_{[i]}}} \right|}^2}}  \le {{\cal T}_{\rm DFT}}\right\},
\end{equation}
\begin{equation}\label{eq:gDft}
{\hat g_{k,i}}={\sqrt{\hat{P_k}}}{\tilde g_{[i]}^*}, i=1,\ldots,\hat P_k
\end{equation}
and
\begin{equation}\label{eq:thetaDft}
{\hat \theta _{k,i}} = \arcsin\{{{2 {I_i}} \mathord{\left/ {\vphantom {{2\pi {I_i}} {M\zeta }}} \right. \kern-\nulldelimiterspace} {M\zeta }}\}, i=1,\ldots,\hat P_k
\end{equation}
respectively, where $\mathcal{T}_{\rm DFT}$ is the energy threshold and ${I_1}, \cdots ,{I_{N_{\rm peak}}}$ are the original indexes of ${\tilde g_{[1]}}, \cdots ,{\tilde g_{[N_{\rm peak}]}}$ in ${{\mathbf{g}}_{\rm{DFT}}}$. Generally, the smaller $\mathcal{T}_{\rm DFT}$, the more paths we will obtain.
\item {\em Downlink channel recovery}: Based on (\ref{eq:pDft}), (\ref{eq:gDft}), (\ref{eq:thetaDft}) and (\ref{eq:hkd}), we acquire the downlink channel estimate at the BS for user $k$ as
\begin{equation}\label{eq:hdDft}
{\mathbf{\tilde h}}_k^{\rm{D}} = \sqrt {\frac{M}{\hat P_k}} \sum\limits_{i = 1}^{\hat P_k} {{{\hat g}_{k,i}}{{\mathbf{a}}_{\rm{D}}}({{\hat \theta }_{k,i}})}.
\end{equation}

\end{enumerate}

In the \emph{spatial matched filtering} step, we scan the space with an oversampling factor $\zeta$ to obtain a better path estimation accuracy. However, this may result in many fake large path gains in the subsequent detection step due to the limited array aperture. To solve this problem, we find the peak elements of ${{\bf{g}}_{{\rm{DFT}}}}$ first and then sort them to detect the true paths in the {\emph{path detection}} step. The energy threshold in the {\emph{path detection}} step is set as
\begin{equation}\label{eq:thresholdDFT}
{{\cal T}_{\rm DFT}}=N/\rho_{\tau},
\end{equation}
whose detailed derivation has been demonstrated in Appendix B. ${{\cal T}_{\rm DFT}}$ in \eqref{eq:thresholdDFT} indicates that the energy threshold for path detection increases with the number of receive antennas but decreases with the pilot transmit SNR.

Note that the proposed DFT-based channel transfer algorithm can be used for $K$ users in parallel to obtain their downlink channel estimation. Regardless of the specific antenna selection methods, this algorithm can also be easily extended to UPA by constructing an appropriate ${{{\mathbf{F}}_{M\zeta }}}$ for UPA.
\begin{algorithm}[t]
\small
	\caption{DFT-based Channel Transfer Algorithm}\label{Alogrithm1}
	\begin{algorithmic}[1]
\REQUIRE  ${\mathbf{\tilde h}}_{k,{\rm A}}^{\rm{U}}$, ${{\mathbf{\tilde h}}_{k,{\rm S}}^{\rm{U}}}={\mathbf{0}}_{M \times 1}$, ${{{\mathbf{F}}_{M\zeta }}}$, ${{\bf{a}}_{\rm{D}}}(\cdot)$, $\hat P_k=1$, $\mathcal{T}_{\rm DFT}$, $M$, $N$, $\mathcal{A}$. \\	
\STATE set ${\left[{{\mathbf{\tilde h}}_{k,{\rm S}}^{\rm U}} \right]_{\mathcal{A}}} = {\mathbf{\tilde h}}_{k,{\rm A}}^{\rm U}$;
\STATE ${{\bf{g}}_{{\rm{DFT}}}} = \frac{{1 }}{ N}{{\bf{F}}_{M\zeta }}{\left[ {{{\left( {{\bf{\tilde h}}_{k,{\rm{S}}}^{\rm{U}}} \right)}^H},{\bf{0}}_{M(\zeta  - 1)}^T} \right]^T}$;
\STATE $\left([{\tilde g_{1}}, \cdots ,{\tilde g_{N_{\rm peak}}}],[{I_1}, \cdots ,{I_{N_{\rm peak}}}]\right)=\text{findpeaks}({\mathbf{g}}_{\rm{DFT}})$;
\STATE $\left([{\tilde g_{[1]}}, \cdots ,{\tilde g_{[N_{\rm peak}]}}]\right)=\text{sort}([{\tilde g_{1}}, \cdots ,{\tilde g_{N_{\rm peak}}}],\text{`descend'})$;
\WHILE {${\left\| {{\bf{\tilde h}}_{k,{\rm{A}}}^{\rm{U}}} \right\|^2} - \sum\limits_{i = 1}^{{{\hat P}_k}} {{{N\left| {{{\tilde g}_{[i]}}} \right|}^2}}  \le {\cal T}_{\rm DFT}$}
		\STATE ${\hat g_{k,{\hat P_k}}}={\sqrt{\hat{P_k}}}{\tilde g_{[\hat P_k]}^*}$;
		\STATE ${\hat \theta _{k,{\hat P_k}}} = \arcsin\{{2 {I_{\hat P_k}}} \mathord{\left/ {\vphantom {{2\pi {I_{\hat P_k}}} {M\zeta }}} \right.	\kern-\nulldelimiterspace} {M\zeta }\}$;
		\STATE $\hat P_k = \hat P_k +1$;
		\ENDWHILE
		\STATE ${\mathbf{\tilde h}}_k^{\rm{D}} = \sqrt {\frac{M}{\hat P_k}} \sum\limits_{i = 1}^{\hat P_k} {{{\hat g}_{k,i}}{{\mathbf{a}}_{\rm{D}}}({{\hat \theta }_{k,i}})}$;
\ENSURE ${\mathbf{\tilde h}}_k^{\rm{D}}$.
	\end{algorithmic}
\end{algorithm}
Additionally, since there is no iteration or matrix inversion calculation, the DFT-based channel transfer algorithm has low computational complexity. The dominant computational complexity of the DFT-based algorithm is resulted by the spatial matched filtering step, which, however, can also be efficiently executed through a fast Fourier transform (FFT). Nevertheless, the performance of the DFT-based algorithm is constrained by the angle mismatch between ${\mathbf{\tilde h}}_{k,{\rm S}}^{\rm U}$ and the DFT columns and the element absence of ${\mathbf{\tilde h}}_{k,{\rm S}}^{\rm U}$, especially when the number of receive antennas is small or there are multiple paths in the propagation channel. Path detection error may occur when these paths are close to each other.

\subsection{mNOMP-based Channel Transfer}
To overcome the path detection error problem and improve the estimation accuracy, we develop another channel transfer algorithm based on Newtonized orthogonal matching pursuit (NOMP).
From (\ref{eq:hdl2}), the uplink-to-downlink channel transfer problem is similar to that of the line spectrum estimation problem \cite{Mamandipoor16Newtonized} but with the absence of elements. Therefore, we propose to modify the NOMP algorithm to restore the downlink channel information. By enhancing the NOMP algorithm with the capability of dealing with the absence of majority elements, the obtained mNOMP algorithm can well solve the uplink-to-downlink channel transfer problem. Furthermore, we also derive an energy threshold for the termination condition step of the mNOMP-based algorithm. By considering the residual noise power into the channel estimates, the proposed mNOMP algorithm can refine the recovered downlink channel estimates as a noise filter. The algorithm is summarized in {\bf Algorithm\,\ref{Alogrithm2}} and the detailed procedure is listed as follows.

\noindent (i) {\em Element completion and initialization}: Pad ${\mathbf{\tilde h}}_{k,{\rm A}}^{\rm U} \in {\mathbb{C}^{N \times 1}}$ with $M-N$ zeros to obtain ${\mathbf{\tilde h}}_{k,{\rm S}}^{\rm U}$ and ${\left[ {{\mathbf{\tilde h}}_{k,{\rm S}}^{\rm U}} \right]_{\mathcal{A}}} = {\mathbf{\tilde h}}_{k,{\rm A}}^{\rm U}$. Initialize the residual vector ${{\mathbf{y}}_r} = {\mathbf{\tilde h}}_{k,{\rm S}}^{\rm U}$, the complex path gain sets $\mathcal{G}=\mathcal{G}_k = \emptyset $, the AoA angle sets $\Omega=\Omega_k = \emptyset $, and the threshold $\mathcal{T}_{\rm m}$;

\noindent (ii) {\em New path detection}: Detect the path with the maximum gain by performing FFT on the residual vector ${{\mathbf{y}}_r}$ and obtain the complex gain $g_{max}$ and the spatial angle $w_{max}$ \footnote{We define $w_{max} \buildrel \Delta \over = \sin\theta_{max}$ and in this section we use ${\bf{a}}(w)$ to denote the steering vector ${\bf{a}}(\arcsin w)$ for notation simplicity.}, i.e.,
    \begin{align}
    ({g_{\max}},{w_{\max}}) &= \mathop {\max }\limits_{|{g_i}|}  \left\{ g_i|{{\bf{g}}_{{\rm{m}}}}={\frac{1}{{ N }}{{\bf{F}}_{M\zeta }}{{\left[ {{\bf{y}}_r^H,{\bf{0}}_{M(\zeta  - 1)}^T} \right]}^T}} \right\},
    \end{align}
    where $g_i$ denotes the elements of ${{\bf{g}}_{{\rm{m}}}}$.
    Then, we update the residual vector as
    \begin{align}
    {{{\bf{\dot y}}}_r} &= {{\mathbf{y}}_r} - {\sqrt N}{g_{\max }}{{\mathbf{a}}_{\rm S}}({w_{max}}),
    \end{align}
     where ${\left[ {{{\mathbf{a}}_{\rm S}}({w_{max}})} \right]_{\mathcal{A},:}} = {{\mathbf{a}}_{\rm U}}({w_{max}})$, ${\left[ {{{\mathbf{a}}_{\rm S}}({ w_{max}})} \right]_{\mathcal{M}\backslash \mathcal{A},:}} = {\mathbf{0}}$ and $\mathcal{M} = \{ 1,2, \cdots ,M\} $;

\noindent (iii) {\em Termination condition}: ${\left\| {{{\mathbf{\dot y}}_r}} \right\|^2}$ is compared with the threshold $\mathcal{T}_{\rm m}$. When ${\left\| {{{\mathbf{\dot y}}_r}} \right\|^2}<\mathcal{T}_{\rm m}$, $\mathcal{G}_k =\mathcal{G}$, $\Omega_k = \Omega $. The algorithm terminates and outputs the downlink channel estimate for user $k$ as
    \begin{equation}
    {\mathbf{\hat h}}_k^{\rm D} = \sqrt{M} \sum\limits_{i = 1}^{\tilde P_k} {{{\tilde g}_i}{{\mathbf{a}}_{\rm D}}({{\tilde w }_i})},
    \end{equation}
    where ${{\tilde g}_i} \in \mathcal{G}_k$, ${{\tilde w }_i}\in \Omega_k$. The estimated number of paths for user $k$ is $\tilde P_k={\left| \mathcal{G}_k  \right|}$;
    When ${\left\| {{{\mathbf{\dot y}}_r}} \right\|^2}\geq\mathcal{T}_{\rm m}$, proceed to the next step;

\noindent (iv) {\em Local gain and angle refinement}: To alleviate the estimation error due to the on-grid angular constraint in the new path detection step, we perform the Newton updates for $ g_{max}$ and $ w_{max}$ with ${{\mathbf{\dot y}}_r}$ by $R_s$ times to obtain optimized values, i.e., for each update we first have
\begin{equation}\label{eq:w_update}
{\tilde w_{\max }} = {w_{\max }} - {{J'_w(w_{max})} \mathord{\left/ {\vphantom {{J'_w(w_{max})} {J^{''}_w(w_{max})}}} \right. \kern-\nulldelimiterspace} {J^{''}_w(w_{max})}},
\end{equation}
\begin{equation}\label{eq:g_update}
{\tilde g_{\max }} = \frac{{{{\left( {{{\bf{a}}_{\rm{S}}}({{\tilde w}_{\max }})} \right)}^H}}}{{{{\left\| {{{\bf{a}}_{\rm{S}}}({{\tilde w}_{\max }})} \right\|}^2}}}\left( {{{{\bf{\dot y}}}_r} + \sqrt N {g_{\max }}{{\bf{a}}_{\rm{S}}}({w_{\max }})} \right),
\end{equation}
where
\begin{equation}
J'_w(w) =  - 2\operatorname{Re} \left\{ {\sqrt N}{{g_{\max }}{\mathbf{\dot y}}_r^H\frac{{\partial {{\mathbf{a}}_{\rm S}}(w )}}{{\partial w }}} \right\},
\end{equation}
and
\begin{align}\label{eq:Jdouble_update}
J^{''}_w(w)=&  - 2\operatorname{Re} \left\{ {\sqrt N}{{g_{\max }}{\mathbf{\dot y}}_r^H\frac{{{\partial ^2}{{\mathbf{a}}_{\rm S}}(w)}}{{\partial {w ^2}}}} \right\} \nonumber \\
&+ 2N{\left| {{g_{\max }}} \right|^2}{\left( {\frac{{\partial {{\mathbf{a}}_{\rm S}}(w )}}{{\partial w }}} \right)^H}\frac{{\partial {{\mathbf{a}}_{\rm S}}(w )}}{{\partial w }},
\end{align}
and then update
\begin{equation}\label{eq:ydot_update1}
{{{\bf{\dot y}}}_r^{\text{temp}}}={{\mathbf{\dot y}}_r} + {\sqrt N}{g_{\max }}{{\mathbf{a}}_{\rm S}}({w _{\max }}) - {\sqrt N}{\tilde g_{\max }}{{\mathbf{a}}_{\rm S}}({\tilde w _{\max }}),
\end{equation}
and
\begin{equation}\label{eq:ydot_update2}
{{{\bf{\dot y}}}_r}={{\mathbf{\dot y}}_r^{\text{temp}}}.
\end{equation}
Note that we update $ w_{max}$ in \eqref{eq:w_update}, $ g_{max}$ in \eqref{eq:g_update} and ${{{\bf{\dot y}}}_r}$ in \eqref{eq:ydot_update2} only when ${J^{''}_w(w_{max})}>0$ and ${\left\| {{{\bf{\dot y}}}_r^{\text{temp}}} \right\|^2} \le {\left\| {{{\bf{\dot y}}}_r} \right\|^2}$.

\noindent (v) {\em Global refinement}: Let ${{{\bf{\ddot y}}}_r} = {{\mathbf{\dot y}}_r}$ and update $\mathcal{G} = \{ \mathcal{G},{\tilde g_{\max }}\} $ and $\Omega  = \{ \Omega ,{\tilde w _{\max }}\} $.
   Then, for each $\tilde w \in \Omega$, cyclically perform the local gain and angle refinement step by $R_c$ times to update the angle estimates in $\Omega$;

\noindent (vi) {\em Global gain corrections}: Based on the $\Omega$ from the global refinement, update all the estimated complex gains in $\mathcal{G}$ by means of the LS method:\\
Let ${\mathbf{A}} = \left[ {{{\mathbf{a}}_{\rm S}}({{\tilde \theta }_1}), \ldots ,{{\mathbf{a}}_{\rm S}}({{\tilde \theta }_{\left| \Omega  \right|}})} \right]$, then
\begin{equation}
{\mathbf{\tilde g}} = {\left( {{{\mathbf{A}}^H}{\mathbf{A}} + {\sigma ^2}{\mathbf{I}_{\left| \Omega  \right|}}} \right)^{ - 1}}{{\mathbf{A}}^H}{\mathbf{\tilde h}}_{k,{\rm S}}^{\rm U}/{\sqrt N},
\end{equation}
where ${\mathbf{\tilde g}} \triangleq {\left[ {{{\tilde g}_1}, \ldots ,{{\tilde g}_{\left| \mathcal{G}  \right|}}} \right]^T}$, ${\left| \Omega  \right|}={\left| \mathcal{G} \right|}$, and $\sigma^2\mathbf{I}_{\left| \Omega  \right|}$ is introduced for retaining full rank of ${\left( {{{\mathbf{A}}^H}{\mathbf{A}} + {\sigma ^2}{\mathbf{I}_{\left| \Omega  \right|}}} \right)^{ - 1}}$. We set ${\sigma ^2} = {10^{ - 4}}$ in the simulation. Then, after updating
\begin{equation}
{{\mathbf{y}}_r} = {\mathbf{\tilde h}}_{k,{\rm S}}^{\rm U} - {\sqrt N}{\mathbf{A\tilde g}},
\end{equation}
we proceed to the new path detection step for the next round of processing until the termination condition is satisfied.

Similar to the DFT-based algorithm, in the proposed mNOMP-based algorithm, we first introduce ${\mathbf{\tilde h}}_{k,{\rm S}}^{\rm U}$ into the element completion step to deal with the element absence. The power variation owing to the absence of the $M-N$ elements is also considered in the algorithm.
In addition, the objective function of the Newton update in the local gain and angle refinement step is
\begin{equation}
\left( {{{\tilde w}_{\max }},{{\tilde g}_{\max }}} \right) = \mathop {\arg \min }\limits_{{w},{g}} {{{\left\| {{{{\bf{\dot y}}}_r} - \sqrt N {g}{{\bf{a}}_{\rm{S}}}({w})} \right\|}^2}},
\end{equation}
and
\begin{equation}
{J_w(w)}={{{\left\| {{{{\bf{\dot y}}}_r} - \sqrt N {g_{\max}}{{\bf{a}}_{\rm{S}}}({w})} \right\|}^2}}.
\end{equation}
$J'_w(w)=\partial {J_w(w)}/\partial w$ and $J^{''}_w(w)=\partial^2 {J_w(w)}/\partial w^2$ are the first and the second-order partial derivatives of ${J_w(w)}$ with respect to $w$, respectively.
The threshold, $\mathcal{T}_{\rm m}$, in the termination condition step is obtained according to the residual noise power. From Appendix B, we have
\begin{equation}
\mathcal{T_{\rm m}}=N/\rho_\tau.
\end{equation}
Note that $\mathcal{T_{\rm m}}$ is different from those in \cite{Mamandipoor16Newtonized,Han19Efficient} where the thresholds are based on the false alarm rate.
Numerical results in Section VI reveal that the proposed mNOMP-based algorithm achieves great performance under both LOS-dominant and multi-path scenarios. Moreover, the antenna array's topology significantly impacts the performance of the proposed mNOMP-based algorithm.

\subsection{Discussion}
As mentioned before, the dominant computational complexity of the DFT-based algorithm comes from \eqref{eq:dftcalculate} in the spatial matched filtering step. Since \eqref{eq:dftcalculate} can be calculated via FFT, the computational complexity of the DFT-based algorithm is $\mathcal{O}({M\zeta }\log{M\zeta })$. As for the mNOMP-based algorithm, the computational complexity mainly comes from the new path detection step, the global refinement step, and the gain correction step, whose computational complexity are $\mathcal{O}(\tilde P_k{M\zeta }\log{M\zeta })$, $\mathcal{O}(M{\tilde P_k^2}R_c R_s)$, and $\mathcal{O}({\tilde P_k^4}+M{\tilde P_k^3})$, respectively. Hence, the computational complexity of the mNOMP-based algorithm is $\mathcal{O}(\tilde P_k{M\zeta }\log{M\zeta }+M{\tilde P_k^2}R_c R_s+M{\tilde P_k^3}+{\tilde P_k^4})$.

Although the mNOMP-based algorithm has much higher computational complexity than the DFT-based algorithm, we always have $\tilde P_k <<M$ and $R_c, R_s <<M$.
In addition, as we emphasized in step (iv) of the mNOMP-based algorithm, the updates of $ w$, $ g$ and ${{{\bf{\dot y}}}_r}$ are only performed when ${J^{''}_w(w)}>0$ and ${\left\| {{{\bf{\dot y}}}_r^{\text{temp}}} \right\|^2} \le {\left\| {{{\bf{\dot y}}}_r} \right\|^2}$. Therefore, at each iteration we only retain the detected new path (i.e., $w_{\max }$ and $g_{\max }$) that decreases the residual signal energy. Furthermore, since the LS update in step (vi) of the mNOMP-based algorithm can also only lead to a residual signal energy reduction \cite{Mamandipoor16Newtonized}, the convergence of the mNOMP-based algorithm is guaranteed.

\begin{algorithm}[h]
\small
	\caption{mNOMP-based Channel Transfer Algorithm}\label{Alogrithm2}
	\begin{algorithmic}[1]
		\REQUIRE  ${{\mathbf{y}}_r} = {\mathbf{\tilde h}}_{k,{\rm S}}^{\rm U}$, ${{\mathbf{a}}_{\rm S}}(\cdot)$, ${{{\mathbf{F}}_{M\zeta }}}$, $\mathcal{G},\mathcal{G}_k = \emptyset $, $\Omega, \Omega_k = \emptyset $, $R_c$, $R_s$, $\mathcal{T}_{\rm m}$, $M$, $N$, $\mathcal{A}$. \\
\WHILE {${\left\| {{{\mathbf{y}}_r}} \right\|^2}\geq\mathcal{T}_{\rm m}$}
        \STATE $({g_{\max}},{w_{\max}}) = \mathop {\max }\limits_{|{g_i}|}  \left\{ g_i|{{\bf{g}}_{{\rm{m}}}}={\frac{1}{{ N }}{{\bf{F}}_{M\zeta }}{{\left[ {{\bf{y}}_r^H,{\bf{0}}_{M(\zeta  - 1)}^T} \right]}^T}} \right\}$;
       \STATE ${{\mathbf{\dot y}}_r} = {{\mathbf{y}}_r} - {\sqrt N}{g_{\max }}{{\mathbf{a}}_{\rm S}}({w _{\max }})$;
       \STATE Refine $g_{\max}$ and $w_{\max}$ with ${{\mathbf{\dot y}}_r}$ by $R_s$ times via \eqref{eq:w_update}-\eqref{eq:ydot_update2}.
       \STATE Let ${{\mathbf{\ddot y}}_r} = {{\mathbf{\dot y}}_r}$, $\mathcal{G} = \{ \mathcal{G},{\tilde g_{\max }}\} $, $\Omega  = \{ \Omega ,{\tilde w _{\max }}\} $;
		
		\FOR{$i=1:R_c$}
			\FOR{each $\tilde w \in \Omega$ and $\tilde g \in \mathcal{G}$}
                \STATE   Refine $\tilde w$ and $\tilde g$ with ${{\mathbf{\ddot y}}_r}$ by $R_s$ times via \eqref{eq:w_update}-\eqref{eq:ydot_update2}.
		    \ENDFOR	
		\ENDFOR
		\STATE ${\mathbf{A}} = [ {{{\mathbf{a}}_{\rm S}}({{\tilde w }_1}), \ldots ,{{\mathbf{a}}_{\rm S}}({{\tilde w }_{\left| \Omega  \right|}})} ]$;\\
        \STATE ${\mathbf{\tilde g}} = {\left( {{{\mathbf{A}}^H}{\mathbf{A}} + {\sigma ^2}{\mathbf{I}}} \right)^{ - 1}}{{\mathbf{A}}^H}{\mathbf{\tilde h}}_{k,{\rm S}}^{\rm U}/{\sqrt N}$;\\
        \STATE ${{\mathbf{y}}_r} = {\mathbf{\tilde h}}_{k,{\rm S}}^{\rm U} - {\sqrt N}{\mathbf{A\tilde g}}$;\\
		\ENDWHILE
\STATE $\mathcal{G}_k =\mathcal{G}$, $\Omega_k = \Omega $, $\tilde P_k={\left| \mathcal{G}_k  \right|}$;
\STATE ${\mathbf{\hat h}}_k^{\rm D} = \sqrt {{M}} \sum\limits_{i = 1}^{\tilde P_k} {{{\tilde g}_i}{{\mathbf{a}}_{\rm D}}({{\tilde w}_i})}$;
		\ENSURE $\mathcal{G}_k, \Omega_k, {\mathbf{\hat h}}_k^{\rm D}$.
	\end{algorithmic}
\end{algorithm}

\section{Cost and Energy Comparison}
In this section, we first provide the cost and the energy consumption models for the asymmetrical transceiver-based massive MIMO systems and then the cost and energy consumption comparisons with other conventional symmetrical transceiver-based massive MIMO systems, e.g., the full digital massive MIMO systems and the hybrid massive MIMO systems, are performed.

\begin{table*}[th]
\caption{The cost and power consumption of hardware components.}\label{tab:parameters}
\scriptsize
\centering
  \begin{tabular}{m{1.4cm}m{0.55cm}m{0.85cm}<{\centering}m{0.55cm}<{\centering}m{0.65cm}<{\centering}m{0.65cm}<{\centering}m{0.65cm}m{0.65cm}<{\centering}m{0.75cm}m{0.75cm}m{0.5cm}b{0.5cm}<{\centering}}
    \toprule[1.2pt]
    {Component} & {PA}& {PA driver}& {LNA}& {Switch} & {Mixer} & {LO amp.} & {Phase shifter}& {IF Tx chain}& {IF Rx chain} & {DAC} & {ADC}\\[3pt]
    \hline
    {Reference Cost (USD)} & $50$& $30$& $27$& $27$& $24$ & $30$& {170}& {140}& {140} & {55} & {451}\\[3pt]
    \rowcolor{mygray}
    {Power Consump. (W)} & $3.68$ & $0.85$ & $0.33$ & $0.1$ &$0$ & $0.6$&{0}& {1.75}& {1.25} & {2.07} & {2.82}\\[3pt]
    \bottomrule[1.0pt]
    \hline
  \end{tabular}
\end{table*}
Table\,\ref{tab:parameters} presents a reference cost and power consumption table of the hardware components that are used to implement a massive MIMO testbed operating at $28$\,GHz with $500$\,MHz bandwidth \cite{Yang18Digital}.
Then, based on the block diagram of the asymmetrical transceiver presented in Fig.\,\ref{Fig:system_architecture} and the architectures of the hybrid transceivers provided in \cite{heath2016overview}, we have
\begin{align}
  {C_{{\text{ADBN}}}} &= M\big( {{C_{{\text{PA}}}} + {C_{{\text{PA driver}}}} + {C_{{\text{Mixer}}}} + {C_{{\text{LO amp}}}} + {C_{{\text{IF Tx chain}}}} }  \nonumber \\
                &\quad + {C_{{\text{DAC}}}}\big)+ N\left( {{C_{{\text{LNA}}}} + {C_{{\text{Switch}}}} + {C_{{\text{IF Rx chain}}}} + {C_{{\text{ADC}}}}} \right),
\end{align}
\begin{align}
{C_{{\text{DBM}}}}& = M( {C_{{\text{PA}}}} + {C_{{\text{PA driver}}}} + {C_{{\text{LNA}}}} + {C_{{\text{Switch}}}} + {C_{{\text{Mixer}}}} \nonumber \\
&\quad + {C_{{\text{LO amp}}}} + {C_{{\text{IF Tx chain}}}}+ {C_{{\text{IF Rx chain}}}} + {C_{{\text{ADC}}}} + {C_{{\text{DAC}}}} ),
\end{align}
\begin{align}
{C_{{\text{HBFN}}}} &= M\left( {{C_{{\text{PA}}}} + {C_{{\text{PA driver}}}} + {C_{{\text{LNA}}}} + 2{C_{{\text{Switch}}}}} \right)  \nonumber \\
                &\quad + MN{C_{{\text{Phase shifter}}}}+ N\big( {C_{{\text{Mixer}}}} + {C_{{\text{LO amp}}}}  \nonumber \\
                &\quad + {C_{{\text{IF Tx chain}}}}+ {C_{{\text{IF Rx chain}}}} + {C_{{\text{ADC}}}} + {C_{{\text{DAC}}}} \big),
\end{align}
\begin{align}
{C_{{\text{HBSN}}}} &= M\left( {{C_{{\text{PA}}}} + {C_{{\text{PA driver}}}} + {C_{{\text{LNA}}}} + 2{C_{{\text{Switch}}}} + {C_{{\text{Phase shifter}}}}} \right) \nonumber \\
                &\quad + N\big( {C_{{\text{Mixer}}}} + {C_{{\text{LO amp}}}} + {C_{{\text{IF Tx chain}}}} + {C_{{\text{IF Rx chain}}}}   \nonumber \\
                &\quad + {C_{{\text{ADC}}}} + {C_{{\text{DAC}}}}\big),
\end{align}
where ${C_{{\text{ADBN}}}}$, ${C_{{\text{DBM}}}}$, ${C_{{\text{HBFN}}}}$, and ${C_{{\text{HBSN}}}}$ represent the cost of the asymmetrical transceiver-based, the conventional full digital symmetrical transceiver-based, the full-connected hybrid transceiver-based, and the subarray hybrid transceiver-based BSs, respectively.\footnote{For the two hybrid transceivers, conventional symmetrical architectures are employed and $N$ denotes the number of RF chains.} The $C_{\{\cdot\}}$ in the right-hand side denotes the cost of the corresponding hardware components.
Consequently, combined with Table\,\ref{tab:parameters}, we obtain
\begin{equation}\label{eq:costComparison}
{C_{{\text{HBFN}}}} > {C_{{\text{DBM}}}} > {C_{{\text{HBSN}}}} > {C_{{\text{ADBN}}}},
\end{equation}
when $M=128$ and $N=16$.
The above inequality indicates that the proposed asymmetrical transceiver achieves the lowest cost while the full-connected hybrid transceiver achieves the highest. This is because, on one hand, fewer ADCs are needed for the asymmetrical transceivers when compared with the full digital symmetrical transceivers; On the other hand, the phase shifters, which are generally utilized in hybrid transceivers (e.g. $MN$ phase shifters are needed for the full-connected hybrid transceiver), are in fact expensive.

Similar to the cost model, we also provide the energy consumption model of the asymmetrical transceiver-based BS as
\begin{align}
  {P_{{\text{ADBN}}}} &= (1 - \varepsilon )M\big( {P_{{\text{PA}}}} + {P_{{\text{PA driver}}}} + {P_{{\text{Mixer}}}} + {P_{{\text{LO amp}}}}  \nonumber \\
                & \quad + {P_{{\text{IF Tx chain}}}} + {P_{{\text{DAC}}}} \big)+ \varepsilon N\big( {P_{{\text{LNA}}}} + {P_{{\text{Switch}}}} + {P_{{\text{Mixer}}}}\nonumber \\
                 &\quad+ {P_{{\text{LO amp}}}} + {P_{{\text{IF Rx chain}}}} + {P_{{\text{ADC}}}} \big),
\end{align}
where $P_{\{\cdot\}}$ in the right hand side denotes the power of the corresponding hardware components; $\varepsilon$ is the ratio of the slots assigned for the uplink to that of the downlink and we set $\varepsilon =1/3$.
Thus, we can also obtain ${P_{{\text{DBM}}}}$, ${P_{{\text{HBFN}}}}$ and ${P_{{\text{HBSN}}}}$, and
\begin{equation}\label{eq:powerComparison}
{P_{{\text{DBM}}}} >  {P_{{\text{ADBN}}}}>{P_{{\text{HBSN}}}}= {P_{{\text{HBFN}}}}.
\end{equation}
This result comes from the fact that the power consumption of phase shifters in the hybrid transceivers are nearly zero and there are less power-hungry ADCs and IF Rx chains in the asymmetrical transceivers.
Moreover, we define the energy efficiency as
\begin{equation}\label{eq:EE}
{\text{EE}} = \frac{{\varepsilon {{\eta _{{\text{SE}}}^{\text{U}}}} + (1 - \varepsilon ){{\eta _{{\text{SE}}}^{\text{D}}}}}}{{{P_{{\text{BS}}}}}} \times {\text{Bandwidth}},
\end{equation}
where ${P_{{\text{BS}}}}$ is one of ${P_{{\text{ADBN}}}}$, ${P_{{\text{DBM}}}}$, ${P_{{\text{HBFN}}}}$ and ${P_{{\text{HBSN}}}}$.
Due to the channel model adopted in (\ref{eq:hku}) and (\ref{eq:hkd}), it is difficult to obtain closed-form expressions for the EE above. Hence, we evaluate the EEs of different transceivers via the Monte-Carlo simulation.
The simulation results in Section VI indicate that, when the numbers of users served at downlink and uplink are the same, the hybrid transceivers achieve a better EE than the asymmetrical transceiver. Nevertheless, thanks to the employed large number of transmit RF chains, the asymmetrical transceivers can simultaneously support more data streams at downlink than that of the uplink. Hence, the downlink EE of the asymmetrical transceivers can be further enhanced when unequal number of users are served at uplink and downlink.

\section{Numerical Results}
In this section, we first evaluate the beam pattern of different receive antenna arrays and uplink SNR loss. Then, the performance of the proposed uplink-to-downlink channel transfer algorithms is investigated.
After that, the uplink and downlink SEs and the corresponding EEs are presented to demonstrate the system performance and the superiority of the proposed asymmetrical transceiver-based systems in contrast to conventional symmetrical massive MIMO systems.

\subsection{Beam Pattern and Uplink SNR Loss}
\begin{figure}[h]
\centering
\includegraphics[scale= 0.35]{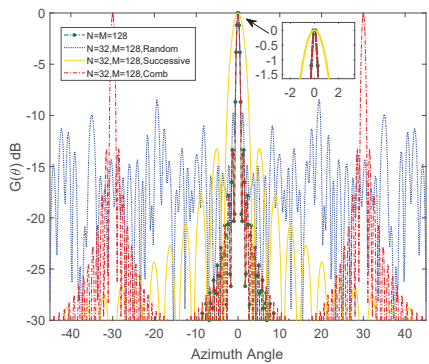}
\caption{ The static directional diagram of different receive antenna arrays when $M=128$ and $N=32$.}\label{fig:staticDirectional}
\end{figure}

We evaluate the angular resolution of different receive antenna arrays via their beam patterns.
Fig.\,\ref{fig:staticDirectional} presents the static directional diagram of the receive antenna array from different antenna selections (i.e., the random antenna selection, the successive antenna selection, and the comb antenna selection).
From Fig.\,\ref{fig:staticDirectional}, a broader beam as well as a lower angular resolution are resulted in the successive antenna array due to the successive antenna element arrangement and consequently a limited effective aperture. In contrast, owing to the distributed antenna element arrangement which in fact spans across the whole antenna array ($M=128$), the receive antenna array with a random-element pattern attains a smaller beamwidth and thus a higher angular resolution \cite{Lo64Mathematical}. Therefore, the random antenna array ($N=32$) acquires the same main lobe width as that of $N=M=128$ and is capable of distinguishing more users or more paths. Despite that, thanks to the large effective array aperture, the array with a comb-element pattern achieves a sharp beam as well. However, because of the larger adjacent element spacing, e.g., $2\lambda$ when $N=32$, grating lobes occur in the comb antenna array and there will be angular ambiguity when interfering paths come from the angles of the grating lobes.

\begin{figure}[htbp]
\centering
\subfigure[The normalized uplink SNR loss.]{
\begin{minipage}[t]{1\linewidth}
\centering
\includegraphics[scale= 0.45]{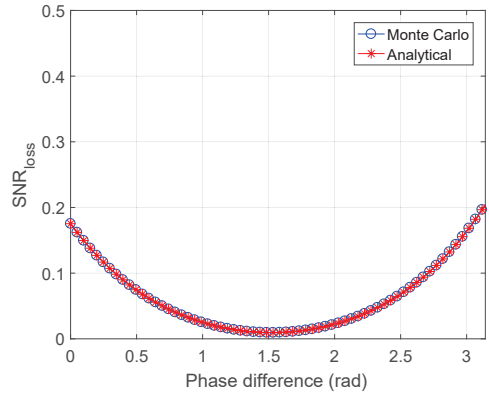}
\end{minipage}%
}%
\vspace{0.1pt}
\subfigure[The resolved paths.]{
\begin{minipage}[t]{1\linewidth}
\centering
\includegraphics[scale= 0.45]{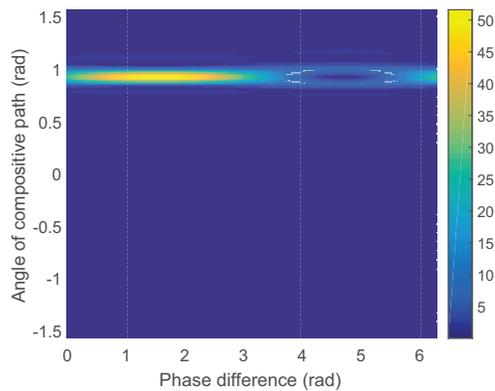}
\end{minipage}%
}%
\centering
\caption{The uplink SNR loss incurred by the degraded angular resolution from successive antenna selection. Two equal-power paths are assumed in the environment. $\Delta_t=\Delta_r=0.45^o$, $\Delta^s_r=3.58^o$, $|\theta_{1}-\theta_{2}|=2.97^o$, and ${\Delta_r=\Delta_t<|\theta_{1}-\theta_{2}| \leq \Delta^s_r}$.}\label{fig:SNRloss}
\end{figure}
Fig.\,\ref{fig:SNRloss} investigates the normalized uplink SNR loss provided in Section III.B. Note that the phase difference is defined as $\phi_2-\phi_1$. From Fig.\,\ref{fig:SNRloss}(a) and (b), it can be observed that, with the increasing phase difference, the $\rm SNR_{loss}$ first decreases and then increases and there is only one dominant path that can be resolved when the phase difference is between $0$ and $\pi$. Nearly $20\%$ SNR loss can be achieved when $\phi_2-\phi_1=\pi$. Besides, Fig.\,\ref{fig:SNRloss}(b) also indicates that two weak paths instead of one dominant path will be resolved even if $|\theta_{1}-\theta_{2}| < \Delta^s_r$ when the phase difference continuously increases (i.e., larger than $\pi$ but less than $2\pi$). This is because $\cos\Gamma<0$ and thus these two paths are in destructive superposition in the anticipated compositive angle, which finally results in two effective paths with a larger angular separation, i.e., $|\theta_{1}'-\theta_{2}'| >\Delta^s_r$.

\subsection{Uplink-to-Downlink Channel Transfer}
\begin{figure}[h]
\centering
\includegraphics[scale= 0.5]{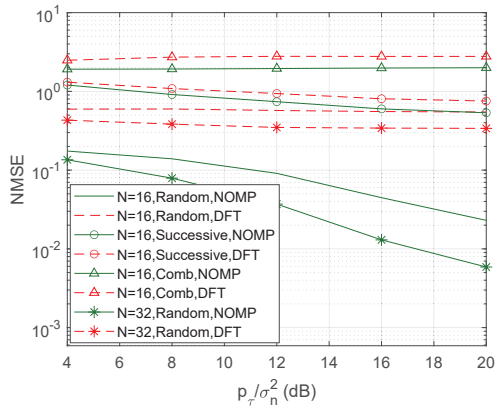}
\caption{ The NMSE of the estimated downlink channel information for user $1$ under LOS path-dominant scenarios. $M=128$ and $R_s=R_c=2$. $P=2$ and the power allocation for paths is $[0.9\quad 0.1]$. $\zeta$ for the DFT- and mNOMP-based algorithms are $8$ and $4$, respectively. }\label{fig:NMSELOS}
\end{figure}

\begin{figure}[h]
\centering
\includegraphics[scale= 0.5]{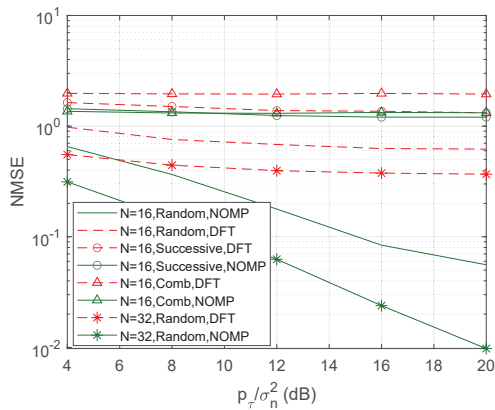}
\caption{ The NMSE of the estimated downlink channel information for user $1$ under multi-path scenarios. $M=128$, $P=3$ and $R_s=R_c=2$.  $\zeta$ for the DFT- and mNOMP-based algorithms are $8$ and $4$, respectively.}\label{fig:NMSENLOS}
\end{figure}

To examine the performance of the uplink-to-downlink channel transfer algorithms, the normalized mean-squared error (NMSE) is utilized as the metric and is defined as
\begin{equation}\label{eq:NMSE}
{\rm NMSE} = \mathbb{E}\left\{ {{{{{\left\| {{\mathbf{\tilde h}}_k^{\rm D} - {\mathbf{h}}_k^{\rm D}} \right\|}^2}}}/{{{{\left\| {{\mathbf{h}}_k^{\rm D}} \right\|}^2}}}} \right\}.
\end{equation}
Note that ${\mathbf{\bar h}}_{k,{\rm A}}^{\rm U}$, i.e., the uplink LMMSE channel estimate, is used as the input of the channel transfer algorithms.
Figs.\,\ref{fig:NMSELOS} and \ref{fig:NMSENLOS} present the NMSE for user 1 under different antenna array topologies for LOS and multi-path scenarios, respectively. The angles of paths are uniformly randomly generated over $[-60^o,60^o]$.
As can be observed from Figs.\,\ref{fig:NMSELOS} and \ref{fig:NMSENLOS}, both the two algorithms benefit from the increasing number of receive antennas and the increasing aperture of array topologies.
With the receive antenna array from the random antenna selection, the mNOMP-based algorithm can acquire nearly $4$\,dB performance improvement when the number of receive antennas $N$ increases from $16$ to $32$ at high SNR regime.
When $N=16$, the random-element pattern achieves the best performance due to its larger array aperture than the successive-element pattern and its better grating lobe suppression than the comb-element pattern. The better grating lobe suppression of the random-element pattern comes from the unequal sized element spacing, which makes the grating lobes replaced by unequal amplitude side lobes all less than the main lobe \cite{King60unequally,Mailloux84array}.  In contrast, the comb-element pattern achieves the worst performance due to the emergence of grating lobes in the visible region, i.e., $[-60^o,60^o]$.
The comb-element pattern can only perform well under extremely narrow angular spread scenarios owing to these grating lobes.
In addition, at the cost of the additional computational complexity, the mNOMP-based algorithm always outperforms the DFT-based algorithm no matter at the LOS scenarios or at the multi-path scenarios.
The DFT-base algorithm is more suitable for fast channel transfer or limited available computational resource scenarios.

\subsection{SE and EE}
\begin{figure}[htbp]
\centering
\subfigure[Uplink.]{
\begin{minipage}[t]{1\linewidth}
\centering
\includegraphics[scale= 0.45]{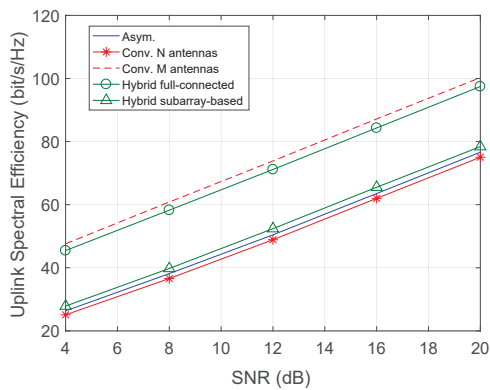}
\end{minipage}%
}%
\vspace{0.1pt}
\subfigure[Downlink.]{
\begin{minipage}[t]{1\linewidth}
\centering
\includegraphics[scale= 0.45]{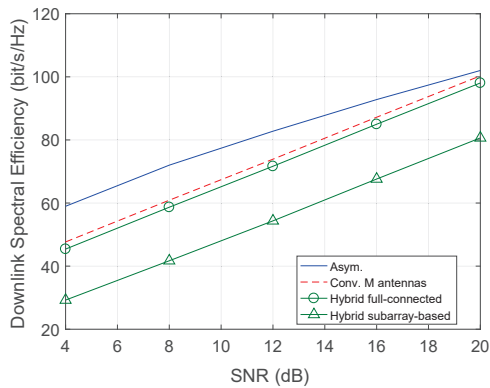}
\end{minipage}%
}%
\centering
\caption{The spectral efficiency comparison.}\label{fig:SEUL}
\end{figure}

Fig.\,\ref{fig:SEUL} shows the uplink and downlink SEs. The receive antennas are selected based on random antenna selection and the mNOMP-based uplink-to-downlink channel transfer algorithm is employed for the asymmetrical transceiver.
For the purpose of comparison, we also provide the SEs of the conventional full digital symmetrical transceiver-based and the hybrid transceiver-based massive MIMO systems. Note that $M=128$, $N=32$, $K=10$, and $P_1=\ldots=P_K=3$ in Fig.\,\ref{fig:SEUL}. LMMSE channel estimation and zero-forcing (ZF) detection are employed for all the systems at the uplink, and ZF precoding is utilized for asymmetrical and conventional full digital systems at the downlink.
As for the hybrid transceiver-based systems, eigenvectors of the uplink channel correlation matrices are used for analog combining at the uplink. While for the downlink precoding, the phased-ZF method proposed by \cite{Liang14Low} is employed for the full-connected hybrid transceiver-based systems and the SIC-based hybrid precoding method proposed by \cite{Gao16Energy} is utilized for the subarray-based hybrid transceiver-based systems, respectively.
Both of these two hybrid precoding methods use the full-dimensional (i.e., $M \times K$) instantaneous channel state information.

\begin{figure}[h]
\centering
\includegraphics[scale= 0.45]{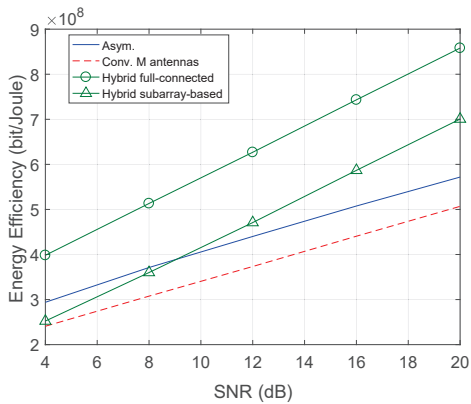}
\caption{ The energy efficiency comparison.}\label{fig:EE}
\end{figure}
Fig.\,\ref{fig:SEUL}(a) shows that the proposed asymmetrical transceiver achieves a similar uplink SE to the subarray-based $M$-antenna hybrid transceiver. Additionally, as we expected, the proposed asymmetrical transceiver achieves a larger uplink SE than the conventional $N$-antenna full digital symmetrical transceiver owing to its flexible receive antenna topology design.
Furthermore, Fig.\,\ref{fig:SEUL}(b) indicates that the proposed asymmetrical transceiver can even outperform the conventional $M$-antenna full digital symmetrical transceiver in terms of the downlink spectral efficiency, especially at the low SNR regime. This is because, by fully exploiting the channel architecture, the proposed mNOMP-based channel transfer algorithm can well recover the key parameters of the downlink channel (e.g., AoA and complex path gains) from the $N$-dimensional uplink LMMSE channel estimates with $N$ receive RF chains.
As a consequence, a higher downlink spectral efficiency can be achieved via the estimates from the proposed mNOMP-based algorithm than that from the pure uplink LMMSE channel estimation with $M(>>N)$ receive RF chains, and thus the downlink transmission capability of the asymmetrical transceiver-based massive MIMO systems can be well maintained.

The energies efficiency of different systems are demonstrated in Fig.\,\ref{fig:EE}. From the figure, the full-connected hybrid transceiver achieves the best energy efficiency. However, due to the employment of a large and complicated phase shifter network, the full-connected hybrid transceiver severely suffers from the tremendous hardware cost and complexity. Additionally, as anticipated, the proposed asymmetrical transceiver achieves a better EE than the conventional $M$-antenna full digital symmetrical transceiver. Moreover, the proposed asymmetrical transceiver also outperforms the subarray-based hybrid transceiver at the low SNR regime. This result mainly comes from the benefits of the full digital architecture of the asymmetrical transceiver and the proposed channel transfer algorithm. Hence, the proposed asymmetrical transceiver-based architecture can be a promising alternative solution for the next generation massive MIMO systems.

\section{Conclusion}

This paper proposed an asymmetrical transceiver-based massive MIMO system by releasing the RF chain pairs and allowing asymmetric transmitter and receiver architecture at the BS. Based on this architecture, the detailed uplink and downlink transmission procedures were investigated and system spectral efficiency and energy efficiency analyses were performed. Besides, to adapt to characteristics of the asymmetrical transceiver-based systems, the uplink-to-downlink channel transfer step was introduced with two channel transfer algorithms, i.e., the DFT-based and the mNOMP-based channel transfer algorithms. The proposed asymmetrical transceiver-based massive MIMO systems could achieve an excellent downlink spectral efficiency and a good system energy efficiency simultaneously. The asymmetrical transceiver-based system architecture provides a promising alternative solution, especially for extra-large scale massive MIMO systems in the future.

\appendices

\section{Proof of Proposition 1}
Since the successive antenna selection is employed, the receive antenna index is set as ${a_n}=n,\forall n=1,\ldots,N$ in (\ref{eq:hku}) for simplicity. In addition, because the single-user scenario and the perfect channel estimation are considered, we have the uplink SNR for user $k$ when these two paths are perfectly resolved as
\begin{align}\label{eq:uplinkSNRUU}
{\rm{SNR}_{\Delta_r}} &= {\rho_u}{{\left\| {{\bf{h}}_{k,{\rm{A}}}^{\rm{U}}} \right\|}^2} ,\nonumber\\
                     & = {\frac{{\rho_u}N}{2}}{\left\| \sum\limits_{i = 1}^{2} {{g_{k,i}}{{\bf{a}}_{\rm{U}}}({\theta _{k,i}})}\right\|}^2.
\end{align}
When there is only one dominant path that has been resolved due to the degraded angular resolution, the uplink SNR becomes
\begin{align}\label{eq:uplinkSNRSS}
&{\rm{SNR}_{\Delta_r^s}} \nonumber \\
%&= {\rho_u} {{{{{\left| {{{\left(  {{\bf{\tilde h}}_{k,{\rm{A}}}^{\rm{U}}} \right)}^H}{\bf{h}}_{k,{\rm{A}}}^{\rm{U}}} \right|}^2}}}/{{ {{\left\|  {{\bf{\tilde h}}_{k,{\rm{A}}}^{\rm{U}}} \right\|}^2}} }} ,\nonumber\\
&= {\rho_u} {{{{{\left| {{{\left(  {\bf{h}}_{k,{\rm{A}}}^{\rm{U,s}} \right)}^H}{\bf{h}}_{k,{\rm{A}}}^{\rm{U}}} \right|}^2}}}/{{ {{\left\|  {\bf{h}}_{k,{\rm{A}}}^{\rm{U,s}} \right\|}^2}} }} ,\nonumber\\
                     & = {{{\rho_u}}}{\left| {{{\left( {g_{k,1}^s{\bf{a}}_{\rm{U}}^s(\theta _{k,1}^s)} \right)}^H}\left( {\sum\limits_{i = 1}^2 {{g_{k,i}}{{\bf{a}}_{\rm{U}}}({\theta _{k,i}})} } \right)} \right|^2}/{\left\| {{g^{{s}}_{k,1}}{{\bf{a}}^s_{\rm{U}}}({\theta^s_{k,1}})}\right\|}^2.
\end{align}
Considering that those two paths are equal power, we assume $g_{k,1}=e^{j\phi_1}$ and $g_{k,2}=e^{j\phi_2}$ in (\ref{eq:hku}) without loss of generality and $|g_{k}^s|=\sqrt 2$ according to the energy conservation law. Then, the uplink SNRs in (\ref{eq:uplinkSNRUU}) and (\ref{eq:uplinkSNRSS}) can be calculated as
\begin{align}\label{eq:uplinkSNRUU2}
{\rm{SNR}_{\Delta_r}} &={\rho_u}N + {\rho_u}\Lambda \cos \left[ {{\phi _1} - {\phi _2} - \frac{{\pi d(N - 1)\Theta }}{\lambda}} \right],
\end{align}
and
\begin{align}\label{eq:uplinkSNRSS2}
&{\rm{SNR}_{\Delta_r^s}} \nonumber \\
  &=\frac{{{\rho_u}(\Lambda _1^2 + \Lambda _2^2)}}{{2N}} + \frac{{\rho_u}{{\Lambda _1}{\Lambda _2}}}{N}\cos \left[ {{\phi _1} - {\phi _2} - \frac{{\pi d(N - 1)\Theta }}{\lambda}} \right],
\end{align}
respectively. Therefore, the normalized SNR loss can be written as
\begin{align}\label{eq:SNRlossDerivation}
&{\rm{SN}}{{\rm{R}}_{{\rm{loss}}}} \nonumber \\
&= \frac{{{\rm{SN{R}}_{{\Delta _{\rm{r}}}}} - {\rm{SN{R}}_{\Delta _{\rm{r}}^{\rm{s}}}}}}{{{\rm{SN{R}}_{{\Delta _{\rm{r}}}}}}},\nonumber\\
& = 1 - \frac{{\Lambda _1^2 + \Lambda _2^2 + 2{\Lambda _1}{\Lambda _2}\cos \left[ {{\phi _1} - {\phi _2} - \pi d(N - 1)\Theta /\lambda} \right]}}{{2{N^2} + 2N\Lambda \cos \left[ {{\phi _1} - {\phi _2} - \pi d(N - 1)\Theta /\lambda} \right]}}.
\end{align}
Substituting $\Gamma  \buildrel \Delta \over = {\phi _1} - {\phi _2} - \pi d(N - 1)\Theta /\lambda$ into (\ref{eq:SNRlossDerivation}), we arrive at (\ref{eq:SNRloss}).

Furthermore, according to (\ref{eq:uplinkSNRUU}), (\ref{eq:uplinkSNRSS}) and the Cauchy-Schwarz inequality, we have ${\rm{SNR}_{\Delta_r^s}}\le{\rm{SNR}_{\Delta_r}}$ and ${\rm{SN{R}}_{{\rm{loss}}}}\ge 0$.
Additionally, since $|\sin (Nx)/\sin (x)| \le N$ and the equality holds when $x=k\pi,k = 0, \pm 1, \pm 2,\ldots$, we have ${\rm{SN{R}}_{{\rm{loss}}}}=0$ when $\Lambda_1=\Lambda_2=\Lambda=N$. This means that when the minimum of ${\rm{SN{R}}_{{\rm{loss}}}}$ is achieved, we have $\pi d\Theta /\lambda  = k\pi$ and $\pi d\Theta_i /\lambda  = k\pi$ for $i=1,2$. Considering that $d=\lambda/2$, $|\Theta|<2$ and $|\Theta_i|<2$ for $i=1,2$, ${\rm{SN{R}}_{{\rm{loss}}}}=0$ indicates $k=0$ and $\theta_{k,1}=\theta_{k,2}$. This result actually contradicts the premise that there are two paths from different angles. Therefore, we have ${\rm{SN{R}}_{{\rm{loss}}}}> 0$.
On the other hand, since $|\cos\Gamma|<1$ and $\Lambda< N$, we have $2N^2+2N\Lambda\cos \Gamma>0$ and ${\Lambda _1^2 + \Lambda _2^2 + 2{\Lambda _1}{\Lambda _2}\cos \Gamma}>(\Lambda_1-\Lambda_2)^2>0$. Thus, the second term of (\ref{eq:SNRlossDerivation}) is larger than zero. Finally, we obtain $0<{\rm{SN{R}}_{{\rm{loss}}}}<1$ and the proof is concluded.

\section{The derivation of ${{\cal T}_{\rm DFT}}$ and $\mathcal{T}_{\rm m}$}
From (\ref{eq:HAuestimatedLS}), we have
\begin{equation}\label{eq:hestimatewithnoise}
{\mathbf{\tilde h}}_{k,{\rm A}}^{\rm U} = {\mathbf{h}}_{k,{\rm A}}^{\rm U}+\frac{1}{\sqrt{\rho_{\tau}}}{\mathbf{\tilde n}}_k,
\end{equation}
and ${\mathbf{\tilde n}}_k\sim\mathcal{CN}({\bf{0}},{{\bf{I}}_N})$ for the LS channel estimation. Assume the number of paths, the complex path gains and the AoAs in the channel transfer algorithms are perfectly recovered. Then for the path detection step in the DFT-based algorithm and the termination condition in the mNOMP-based algorithm, we have
\begin{align}\label{eq:thresholderivation}
\mathbb{E}\left\{{\left\| {{\bf{\tilde h}}_{k,{\rm{A}}}^{\rm{U}}} \right\|^2} - \sum\limits_{i = 1}^{{{\hat P}_k}} {{{N\left| {{{\tilde g}_{[i]}}} \right|}^2}} \right\}
&=\mathbb{E}\left\{{\left\| {{\bf{\tilde h}}_{k,{\rm{A}}}^{\rm{U}}} \right\|^2} - {\left\| {{\bf{ h}}_{k,{\rm{A}}}^{\rm{U}}} \right\|^2} \right\},\nonumber\\
&\mathop  = \limits^{(a)} \mathbb{E}\left\{{\left\| {{{{\bf{\tilde n}}}_k}} \right\|^2}/{\rho _\tau }\right\},\nonumber\\
&=N/{\rho _\tau },
\end{align}
where (a) utilizes the independence between ${\mathbf{h}}_{k,{\rm A}}^{\rm U}$ and ${\mathbf{\tilde n}}_k$.
(\ref{eq:thresholderivation}) indicates that the channel estimate error due to the channel noise is $N/{\rho _\tau }$. Hence, to detect all the paths, we set ${{\cal T}_{\rm DFT}}=\mathcal{T}_{\rm m}=N/{\rho _\tau }$.
Note that from (\ref{eq:HAuestimatedLMMSE2}), it can be observed that (\ref{eq:hestimatewithnoise}) also holds for the LMMSE channel esimation when SNR is high. Therefore, the derived ${{\cal T}_{\rm DFT}}$ and $\mathcal{T}_{\rm m}$ can also be applied for LMMSE channel estimation when $\rho_{\tau}$ is large.

%\section*{Acknowledgment}


\ifCLASSOPTIONcaptionsoff
  \newpage
\fi



%\bibliographystyle{IEEEtran}
%\begin{thebibliography}

\footnotesize
%\bibliographystyle{IEEEtran}
%\bibliography{myreference}

%\begin{biography}[{\includegraphics[width=1in,height=1.25in,clip,keepaspectratio]{mshell}}]{Michael Shell}

%\end{IEEEbiography}


%\newpage



%\vfill



\end{document}